\documentclass[a4paper,aps,pra,twocolumn,showpacs,preprintnumbers,amsmath,amssymb]{revtex4}
\usepackage{bbm}
\usepackage{amsmath}
\usepackage{verbatim}
\usepackage{graphicx}
\bibliography{References}
\begin{document}

\title{Dissipatively driven entanglement of two macroscopic atomic ensembles}

\author{Christine A. Muschik$^1$, Eugene S. Polzik$^{2}$, and J. Ignacio Cirac$^1$}

\affiliation{ $^1$Max-Planck--Institut f\"ur Quantenoptik,
Hans-Kopfermann-Strasse, D-85748 Garching, Germany \\
$^2$ Niels Bohr Institute, Danish Quantum Optics Center Â
QUANTOP, Copenhagen University, Blegdamsvej 17, 2100 Copenhagen
Denmark.}

\begin{abstract}
Up to date, the life time of experimentally demonstrated entangled
states has been limited, due to their fragility under decoherence
and dissipation. Therefore, they are created under strict
isolation conditions. In contrast, new approaches harness the
coupling of the system to the environment, which drives the system
into the desired state. Following these ideas, we present a robust
method for generating steady state entanglement between two
distant atomic ensembles. The proposed scheme relies on the
interaction of the two atomic systems with the common vacuum modes
of the electromagnetic field which act as an engineered
environment. We develop the theoretical framework for two level
systems including dipole-dipole interactions and complement these
results by considering the implementation in multi-level ground
states.
\end{abstract}

\pacs{03.67.Mn, 32.80.Qk}

\maketitle 

\section{Introduction}

Entanglement is the most prominent and distinctive feature of
quantum mechanics. As it plays a central role in fundamental tests
of Quantum Mechanics and in applications in the field of quantum
information science, entanglement has been generated and studied
in various systems. However, experimental generation of entangled
states and their use in quantum information and communication
protocols is hampered by their fragility under decoherence. The
life time of entangled quantum states is typically very short. The
pursuit of the generation of persistent entanglement is not only
of fundamental interest in view of the investigation of entangled
quantum states at long time-scales but also vital for many
applications, such as quantum repeaters \cite{Briegel,DLCZ,
QcomputerReview,Kuzmich,Kimble,Lukin,QuantumInternet,yuan2008,
Kimble08,Pan,InterfaceReview}. On account of this problem, quantum
systems are usually strictly isolated in the endeavor to avoid
their interaction with the environment.

In contrast, we adopt an ostensibly counter-intuitive approach
using dissipation
\cite{PoyatosCiracZoller,Plenio99,Knight2000,Vedral2000,
Braun,myatt2000,Plenio2002,Song2003,KrausCirac,Baumgartner,ManciniWang,Wubs1,KrausZoller2,branderhorst2008,syassen2008,KrausZoller1,Wubs2,FrankWolfCirac,Davidovich,Almut1,Isar,Klemens,Angelakis1,Angelakis2,Rydberg,Marzolino,TemmeWolfVerstraete,Morigi,PrecMeasurement,Wang2010,deValle2010,Kiffner1,Kiffner2,Engineering,barreiro2010}.
Here, the interaction of the system with the environment is
employed such that dissipation drives the system into the desired
state.
More specifically, we propose and analyze a scheme for the
generation of long-lived entanglement between two distant,
mesoscopic ensembles (see also \cite{Scott}). Both atomic samples
are placed in magnetic fields and interact with an environment
consisting of the vacuum modes of the electromagnetic field. A
laser field mediates the coupling of the atomic system to the
environment. The interaction of the system and the bath can be
controlled via laser- and magnetic fields, which allow one to
engineer the coupling in such a way that the unique steady state
of the dissipative evolution is an entangled state.

This dissipative approach has several remarkable advantages. For
example, the scheme performs well starting from an arbitrary
initial state. This feature renders the initialization of the
system in a pure state unnecessary. Most importantly, the
evolution is robust against moderate external noise. Entanglement
is obtained in a steady state. This auspicious property is very
promising in view of the quest for viable, extremely long-lived
entanglement.

We develop a scheme for two-level systems and show that
steady-state entanglement can be generated in the presence of
undesired transitions and fluctuating magnetic fields. We also
include external pump fields and find that surprisingly,
incoherent pumping can be beneficial in a certain parameter range.
These central results are supplemented by a short and very general
study of the implementation in atoms with muli-level ground
states. Since additional dynamics in atoms with multi-level
structure lead to particle losses, a quasi-steady state is
produced. Remarkably, incoherent pumping enables the creation of
entanglement in a true steady state. We investigate the conditions
for the generation of  long-lived entanglement in mesoscopic
multi-level systems, using a simplified model and consider the
realization of the proposed method in $^{133}\text{Cs}$ vapors as
used in \cite{Duan2000,Polzik2001,Squeezing} as specific example.

The paper is organized as follows. The main idea is introduced in
Sec.~\ref{BasicIdeaCentralResults}, which also contains a summary
of the central results. In Sec.~\ref{TwoLevelModel} we derive the
full master equation for two-level atoms and calculate the
evolution of entanglement. Prospects for generating steady-state
entanglement in multi-level systems are discussed in
Sec.~\ref{Alkali Implementation}. Sec.~\ref{Conclusions}
summarizes and concludes the paper.
\begin{figure}[pbt]
\begin{center}
\includegraphics[width=7.5cm]{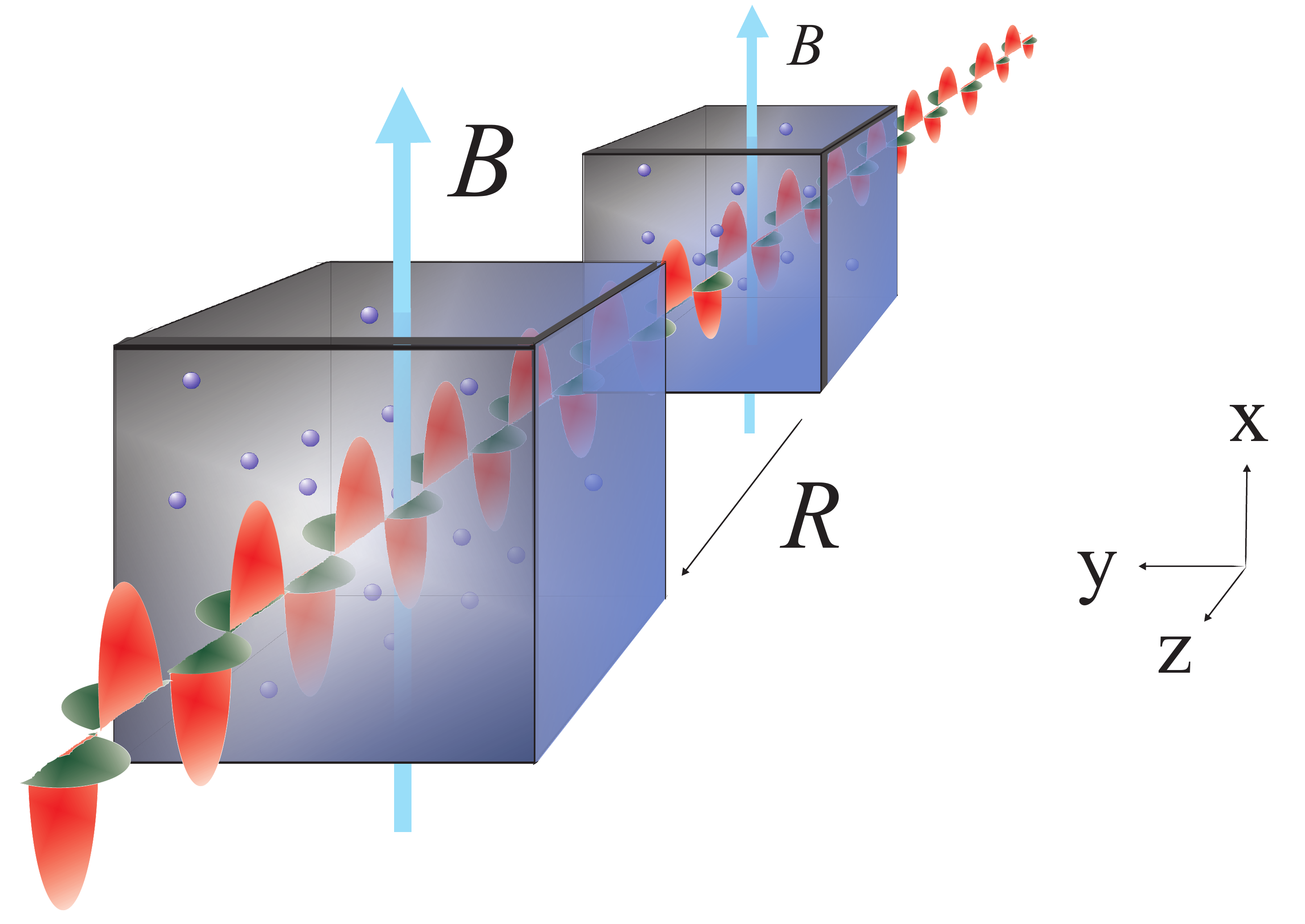}
\caption{(Color online) Setup for creating steady-state
entanglement between two atomic ensembles separated by a distance
$\textbf{R}$. Both ensembles are placed in magnetic fields
$\mathbf{B}$, which are oriented along $\hat{\mathbf{x}}$. A
strong $\hat{\mathbf{y}}$-polarization laser beam (shown in green
color) propagates along $\hat{\mathbf{z}}$ and couples the atomic
system to the environment which consists of the vacuum modes of
the copropagating electromagnetic field in $\hat{\mathbf{x}}$
polarization (depicted in red).  The interaction between atoms and
light modes is illustrated in Fig.~\ref{Figure2}.}\label{Figure1}
\end{center}
\end{figure}
\section{Main idea and central results}\label{BasicIdeaCentralResults}
In the following, we explain the basic idea for generating purely
dissipatively driven entanglement in atomic ensembles. We
introduce a realistic description including noise effects and
discuss the prospects for realizing the proposed scheme
experimentally.

We start by explaining the underlaying concept for two bosonic
modes with annihilation operators $a$ and $b$.
The entangled target state under consideration is a two mode
squeezed (TMS) state $|\Psi_{\text{TMS}}\rangle$, which is
characterized in terms of
\begin{eqnarray*}
\tilde{A} |\Psi_{\text{TMS}}\rangle = \tilde{B}
|\Psi_{\text{TMS}}\rangle =0,
\end{eqnarray*}
where the nonlocal annihilation operators $\tilde{A}$ and
$\tilde{B}$ \cite{AB} are given by
\begin{eqnarray}\label{AB}
    \tilde{A} &=&\mu\  a+\nu \ b^{\dag},\\
    \tilde{B} &=&\mu\  b+\nu \ a^{\dag}.\nonumber
\end{eqnarray}
This equation completely characterizes a particular squeezed state
with squeezing parameter $r$ where $\mu= \cosh(r)$ and
$\nu=\sinh(r)$.
This state can be prepared by means of a dissipative evolution
governed by the master equation
\begin{eqnarray}
d_{t}\rho(t)\!\!&=&\!\!\kappa_{\tilde{A}} \left( \tilde{A} \rho(t)
\tilde{A}^{\dag}-\tilde{A}^{\dag} \tilde{A}
\rho(t)/2-\rho(t)\tilde{A}^{\dag}\tilde{A}/2\right)\nonumber\\
\!\!&+&\!\! \kappa_{\tilde{B}}\left( \tilde{B} \rho(t)
\tilde{B}^{\dag}-\tilde{B}^{\dag} \tilde{B}
\rho(t)/2-\rho(t)\tilde{B}^{\dag}\tilde{B}/2\right)\!,\
\label{MEideal}
\end{eqnarray}
where the coefficients $\kappa_{\tilde{A}}$ and
$\kappa_{\tilde{B}}$ are positive. This time evolution drives the
system into the unique steady state
$\rho_{\infty}=|\Psi_{\text{TMS}}\rangle\langle\Psi_{\text{TMS}}|$,
for $t\rightarrow\infty$ (see App.~\ref{Uniqueness}).
\\Starting from this result, we explain how this concept can be applied for
creating entanglement between two macroscopic atomic ensembles and
how a dissipative evolution of type (\ref{MEideal}) can be
realized.
We consider two atomic ensembles placed in magnetic fields as
shown in Fig.~\ref{Figure1}. Atoms are assumed to possess a
two-level ground state with internal states $|
\!\!\uparrow\rangle$ and $|\!\!\downarrow\rangle$ (see
Fig.~\ref{Figure2}). We aim at creating an entangled state, where
the collective spins of the two samples are correlated in
$\hat{\mathbf{y}}$ and in $\hat{\mathbf{z}}$ direction. The amount
of entanglement generated can be measured by the quantity
\cite{Sanders, Polzik2001}
\begin{eqnarray}
 \xi = \frac{\rm {var} (J_{y,I} + J_{y,II})+ \rm var (J_{z,I} -J_{z,II})}{|\langle J_{x,I}\rangle| + |\langle
 J_{x,II}\rangle|}.\label{xi}
\end{eqnarray}
For separable states $\xi\geq1$.
$J_{x,I}=\sum_{i=1}^{N_{I}}j_{x,I}^{i}$ is the collective spin in
$\hat{\mathbf{x}}$ direction in the first ensemble, where $N_{I}$
is the number of atoms and $j_{x,I}^{i}$ is the
$\hat{\mathbf{x}}$-spin component of the $i$th atom
\cite{FTspins}. Analogous definitions hold for for spins in
$\hat{\mathbf{y}}$ and $\hat{\mathbf{z}}$ direction and for the
collective spin of the second ensemble.
In order to prepare this target state, we use a dissipative
evolution governed by a master equation of type (\ref{MEideal}),
where the nonlocal operators $\tilde{A}$ and $\tilde{B}$ are
replaced by
\begin{eqnarray}\label{AB}
    A&=&\mu J_{I}^{-}+\nu J_{II}^{+}\ ,\\
    B&=&\mu J_{II}^{-}+\nu J_{I}^{+}\ .\nonumber
\end{eqnarray}
$\mu,\nu\in \mathbb{R}$ and $J^\pm_{\text{I/II}}$ are collective
spin operators \cite{FTapproxmation}, with
$J^-=\frac{1}{\sqrt{N}}\sum_{i=1}^{N}|\!\!\uparrow\rangle_{i}\langle
\downarrow\!\!|$ and
$J^+=\frac{1}{\sqrt{N}}\sum_{i=1}^{N}|\!\!\downarrow\rangle_{i}\langle
\uparrow\!\!|$ such that $J_{y}=\frac{1}{2}(J^{+}+J^{-})$ and
$J_{z}=\frac{i}{2}(J^{+}-J^{-})$. Normalization of the operators
$[A,A^{\dag}]=[B,B^{\dag}]=1$ requires $\mu^2-\nu^2=1$.
The light-matter interaction shown in Fig.~\ref{Figure2} gives
rise to the desired master equation. More specifically, after
adiabatic elimination of excited states, the effective ground
state Hamiltonian is of the form $H\propto \int_{\Delta
\omega_{ls}}d\mathbf{k}\left(A
a_\mathbf{k}^{\dag}+A^{\dag}a_\mathbf{k}\right)+\int_{\Delta
\omega_{us}}d\mathbf{k}\left(B
a_\mathbf{k}^{\dag}+B^{\dag}a_\mathbf{k}\right)$, where
$a_{\mathbf{k}}^{\dag}$ is the creation operator for a photon with
wave vector $\mathbf{k}$. The first and second integral cover
narrow bandwiths $\Delta \omega_{ls}$ and $\Delta \omega_{us}$
centered around the lower and upper sideband respectively (see
Sec.~\ref{MasterEquation1}). The modes of the light field are
treated as bath and are therefore traced out. Using the
Born-Markov approximation, one obtains a master equation of
standard Lindblad form (compare Eq.~(\ref{MEideal})). Collective
Lamb shifts can be shown to be negligible in the setting
considered here (see App.~\ref{DerivationME-II}). In the limit
$t\rightarrow\infty$, this evolution drives the system into an
entangled steady state. In the absence of noise and for $N\gg 1$,
\begin{eqnarray*}
\xi_{\infty}^{\text{ideal}}=\left(|\mu|-|\nu|\right)^{2}.
\end{eqnarray*}
Next, we include additional processes such as thermal motion,
undesired atomic transitions and fluctuating magnetic fields as
well as resonant pump fields. Details can be found in
Sec.~\ref{MasterEquation3}. For large particle numbers
$N_{I}=N_{II}=N\gg 1$ and $t\rightarrow\infty$, we find
\begin{eqnarray}
\xi_{\infty}=
 \frac{1}{P_{2,\infty}}\ \frac{\tilde{\Gamma}+d\Gamma
 P_{2,\infty}^2\left(|\mu|-|\nu|\right)^2}{\tilde{\Gamma}+d\Gamma
 P_{2,\infty}},\label{MainResult}
\end{eqnarray}
where $P_{2,\infty}$ is the steady state value of the atomic
polarization
$P_2=\frac{1}{N}\sum_{i=1}^{N}\left(|\!\!\uparrow\rangle_{i}\langle\uparrow\!|-|\!\!\downarrow\rangle_{i}\langle\downarrow\!|\right)$,
$\Gamma$ is the single particle decay rate and $\tilde{\Gamma}$ is
the dephasing rate associated with noise effects. $d$ is the
optical depth of one atomic ensemble.
As shown in Sec.~\ref{CreationOfEntanglement}, the application of
resonant pump fields can be beneficial even though noise is added
by doing so.
Note that for $d\rightarrow\infty$, $\xi_{\infty}\rightarrow
\xi_{\infty}^{\text{ideal}}$. For a large optical depth, the
entangling dynamics is significantly enhanced by collective
effects. In contrast, noise processes are single particle effects
and therefore not amplified by a factor $d$.
Eq.~(\ref{MainResult}) shows that for strong coupling between
atoms and light, entanglement can be generated in a steady state.
This is the main result of this article.\\
\\Intuitively, entanglement is created by virtue of interference of
different processes in the first and second ensemble. As
illustrated in Fig.~\ref{Figure2}, the interaction of light and
atoms in the first ensemble is chosen such that the emission of a
photon in the upper sideband corresponds to a spin flip
$|\!\!\uparrow\rangle \rightarrow |\!\!\downarrow\rangle$.
Similarly, the emission of a photon in the upper sideband involves
a spin flip $|\!\!\downarrow\rangle \rightarrow
|\!\!\uparrow\rangle$ in the second ensemble.
Due to collective effects \cite{InterfaceReview}, light is emitted
in forward direction with high probability, for hot samples with
high optical depth. As spin flips in either ensemble lead to
emission of light into the same spatial mode, both processes are
indistinguishable if a photon is detected. (An analogous argument
holds for photons in the lower sideband.)
In this respect, the setup resembles quantum repeater schemes
\cite{DLCZ,
QcomputerReview,Kuzmich,Kimble,Lukin,QuantumInternet,yuan2008,
Kimble08,Pan}, where collective excitations in two atomic
ensembles are converted to photons, which subsequently interfere
at a 50/50 beamspitter such that entanglement can be created
conditioned on the detection of a photon in one of the two output
ports of the beamsplitter \cite{Analogy}.
Here, no beamsplitter is needed, since both ensembles emit into
the same spatial mode. The most important difference, however,
lies in the fact that our scheme is not conditioned on a specific
measurement outcome. It works deterministically and does not
require a detection of the emitted photon, as the measurement is
performed continuously by the environment.\\
\\The ideas put forward in this work are devised and elaborated for
two-levels systems, but the proposed scheme can also be realized
using atoms with multi-level ground states. The scheme put forward
in this work has been realized recently \cite{ShortPaper} using
alkai atoms. In this experiment, purely dissipatively driven
entanglement between two macroscopic atomic ensembles at room
temperature has been demonstrated yielding an order of magnitude
improvement in the entanglement life time compared to previous
experiments, where entanglement has been generated in this system
using standard methods. In a multi-level setting, the population
in the two-level subsystem is continually reduced due to undesired
transitions to other ground state levels. We take this and other
features of the multilevel structure into account and find that
the continuous reduction of the collective spin leads to the
production of a quasi steady state: the steady state with respect
to the relevant two-level subsystem is superposed by slow
additional dynamics due to the multi-level structure. This can be
counteracted by the application of external pump fields. These
fields add noise to the system and limit therefore the amount of
entanglement that can be generated. However, for samples with high
optical depth, incoherent pumping can render the creation of
steady state entanglement in atoms with multi-level ground states
possible. This is illustrated in Sec.~\ref{MultilevelCalculation}
by considering $^{133}$Cs vapors at room temperature and
experimental parameters close to the values published in
\cite{Polzik2001,Squeezing,ShortPaper}. We take the most
fundamental limitations imposed by undesired radiative processes
into account and estimate that steady state entanglement with
$\xi_{\infty}=0.9$ should be attainable for a moderate optical
depth $d=30$ in the absence of other (implementation-dependent)
sources of noise.
\section{Creation of steady state entanglement in a two-level system}\label{TwoLevelModel}
\begin{figure}[pbt]
\begin{center}
\includegraphics[width=8.5cm]{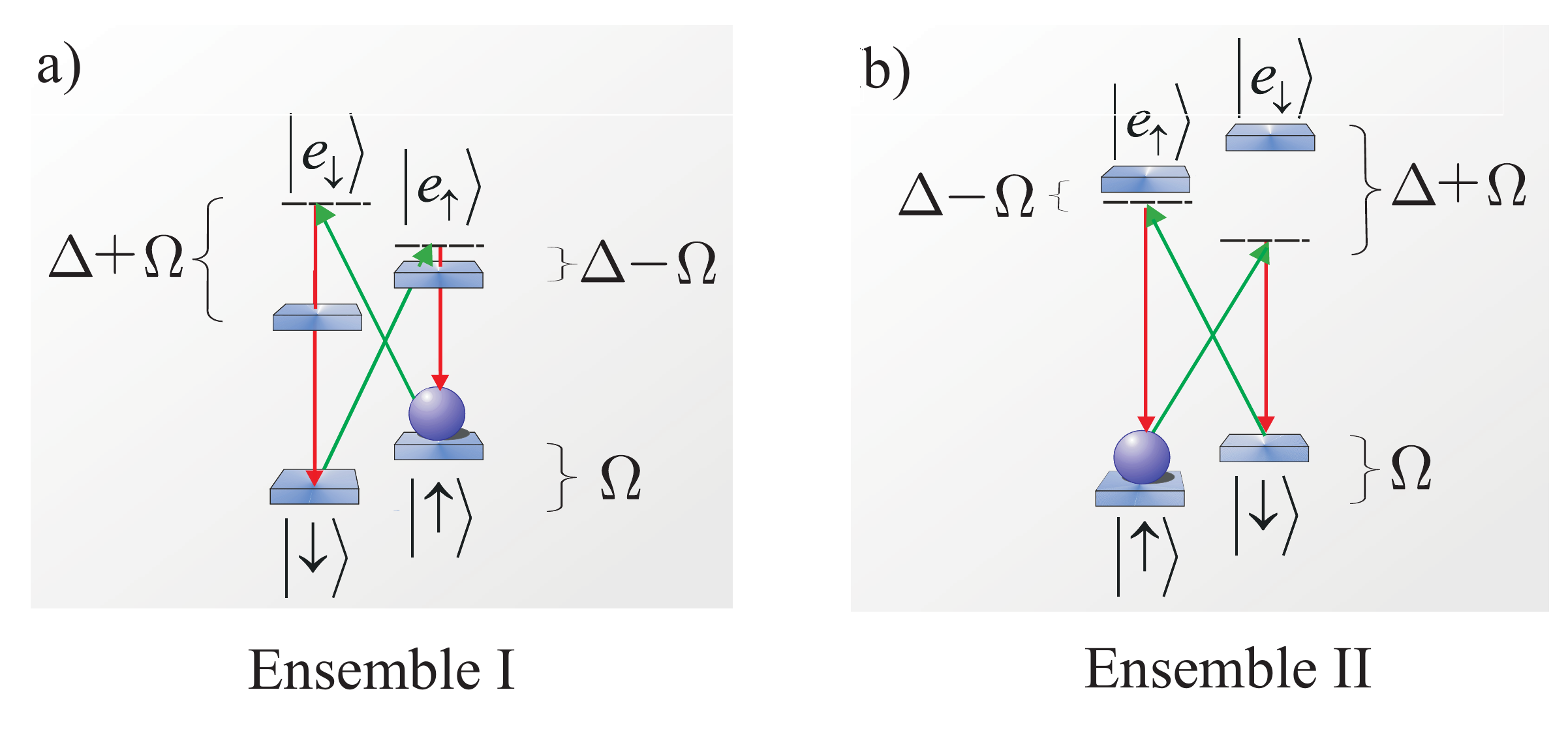}
\caption{(Color online) Atomic level schemes. a) A magnetic field
which is oriented along $\hat{\mathbf{x}}$ (see
Fig.~\ref{Figure1}) causes a Zeeman splitting $\Omega$ of atomic
ground-states levels $|\!\!\uparrow \rangle$ and
$|\!\!\downarrow\rangle$ and defines the quantization axis. A
strong $\hat{\mathbf{y}}$ polarized coherent field with detuning
$\Delta$ drives transitions $|\!\!\!\uparrow\rangle \!\rightarrow
\!|e_{\downarrow}\!\rangle$ and $|\!\!\!\downarrow\rangle
\!\rightarrow\! |e_{\uparrow}\rangle$. Coupling to the vacuum
modes of the electromagnetic field gives rise to transitions
$|e_{\downarrow}\!\rangle \rightarrow |\!\!\downarrow\rangle$ and
$|e_{\uparrow}\rangle \rightarrow |\!\!\uparrow\rangle$. b) A
static electric field is applied to the second ensemble such that
the energy difference between ground and excited states is
enhanced by $2 \Delta$.}\label{Figure2}
\end{center}
\end{figure}
%
%
%
%
As outlined above, light modes act as environment and the
interaction between the system and the reservoir is controlled by
means of laser- and magnetic fields. In
Sec.~\ref{MasterEquation1}, we explain the interaction between
atoms and light in more detail.
The master equation governing the dissipative evolution of the
reduced density matrix of the atomic system $\rho$ is given by
 \begin{eqnarray*}
d_t\rho = {\cal L}_{\rm ent}\rho + {\cal L}_{\rm
 noise}\rho,
 \end{eqnarray*}
where ${\cal L}_{\rm ent}$ and ${\cal L}_{\rm noise}$ are Lindblad
operators. Desired interactions give rise to the entangling
dynamics represented by ${\cal L}_{\rm ent}\rho$. The second term
${\cal L}_{\rm
 noise}\rho$ summarizes all undesired effects.

Below, ${\cal L}_{\rm ent}\rho $ and  ${\cal L}_{\rm noise}\rho$
are determined. To this end, the master equation corresponding to
the light-matter interaction in Figs.~\ref{Figure1}
and~\ref{Figure2} is derived in Sec.~\ref{MasterEquation2},
including undesired radiative processes. Excited states are
adiabatically eliminated such that an effective master equation
for atomic ground states $|\!\!\uparrow\rangle$ and
$|\!\!\downarrow\rangle $ is obtained. In
Sec.~\ref{MasterEquation3}, thermal motion of atoms is taken into
account and additional effects due to pump fields and noise
processes are included.
Based on these results, the amount of entanglement that can be
produced is calculated in Sec.~\ref{CreationOfEntanglement}.
\subsection{Light--matter interaction}\label{MasterEquation1}
%
%
%
In this subsection, we describe the setup for creation of
entanglement between two atomic ensembles and explain how light
and matter interact.\\
%
%
\\We consider the setup shown in Fig.~\ref{Figure1}. A strong
$\hat{\mathbf{y}}$-polarized laser beam propagates along
$\hat{\mathbf{z}}$ and passes two atomic ensembles in a
homogeneous magnetic field, which defines the quantization axis
and is oriented along $\hat{\mathbf{x}}$. Each atomic ensemble
consists of a large number $N$ of hydrogen-like atoms with an
internal level structure as depicted in Fig.~\ref{Figure2}.
The laser field is assumed to cover a very narrow bandwidth $b$
around the central frequency $\omega_{L}$ and to be off-resonant
such that the interaction is well within the dispersive regime and
absorption effects can be neglected. The detuning $|\Delta|$ is
considered to be large compared to the Doppler width
$\delta_{\text{Doppler}}$ and atomic decay rates
$\Gamma_{\text{atomic}}$. Here and in the following,
$\Gamma_{\text{atomic}}$ denotes the largest effective atomic rate
for transitions between ground state levels, including single
particle as well as collective rates (see below). The magnetic
field causes a Zeeman splitting of the atomic ground states
$\Omega$.
The strong $\hat{\mathbf{y}}$-polarized coherent beam is treated
as classical field. With respect to quantization along
$\hat{\mathbf{x}}$, it drives diagonal transitions
$|\!\!\uparrow\rangle\rightarrow|e_{\downarrow}\rangle$,
$|\!\!\downarrow\rangle\rightarrow|e_{\uparrow}\rangle$.
Figs.~\ref{Figure1} and \ref{Figure2} depict only desired
transitions, where photons are scattered into the copropagating
$\hat{\mathbf{x}}$-polarized quantum field in two independent
frequency bands, the upper and the lower sideband, centered around
$\omega_{L}\pm \Omega$.\\
%
%
\\For the realization of the proposed scheme, several setups are
possible. In a simple two-level model, where the Larmor splitting
of excited states equals the Larmor splitting of ground states, a
homogeneous static electric field can be applied to the second
ensemble such that the resulting Stark shift enhances the energy
difference between ground and excited states by $2 \Delta$ as
shown in Fig.~\ref{Figure2}. This yields the following effective
ground state Hamiltonian
\begin{eqnarray*}
H=H_{A}+H_{L}+H_{\text{int}},
\end{eqnarray*}
where excited states have been eliminated under the condition
$|\Delta|\gg
 \Gamma_{\text{atomic}}, \ \delta_{\text{Doppler}}$. Throughout
 the whole paper, the convention $\hbar\equiv1$ is used.
$H_{A}= \Omega \left(J_{x,I}-J_{x,II}\right)$ accounts for the
Zeeman splitting of atoms in the external magnetic field and
$H_{L}=\int dk  \left(\omega_k-\omega_{L}\right)
a_{k}^{\dag}a_{k}$ is the free Hamiltonian of the light field. In
a rotating frame, the interaction Hamiltonian is given by
\begin{eqnarray}
H_{\text{int}}\!\!\!&=&\!\!\!\!\!\int_{\Delta
\omega_{ls}}\!\!\!\!\!\!\!\!\!\! d
\mathbf{k}\sum_{\lambda_{\mathbf{k}}}
\bar{g}(\mathbf{k})\!\!\left(\!\!\mu\!\sum_{i=1}^{N}\sigma_{I,i}e^{i\Delta\mathbf{k}\
\!\!\mathbf{r}_{i}}
\!+\!\nu\!\!\sum_{j=1}^{N}\sigma_{II,j}^{\dag}e^{i\Delta\mathbf{k}\
\!\!\mathbf{r}_{j}}\!\!\right)\!\!a_{\mathbf{k}}^{\dag}\nonumber\\
&+&\!\!\!\!\!\int_{\Delta \omega_{us}}\!\!\!\!\!\!\!\!\!\!d
\mathbf{k}\sum_{\lambda_{\mathbf{k}}} \bar{g}(\mathbf{k})
\!\!\left(\!\!\mu\!\sum_{i=1}^{N}\sigma_{II,i}e^{i\Delta\mathbf{k}\
\!\!\mathbf{r}_{i}}
\!+\!\nu\!\!\sum_{j=1}^{N}\sigma_{I,j}^{\dag}e^{i\Delta\mathbf{k}\
\!\!\mathbf{r}_{j}}\!\!\!\right)\!\!a_{\mathbf{k}}^{\dag} \nonumber\\
&+&H.C.\ ,\label{Hamiltonian}
\end{eqnarray}
where the first and second integral cover narrow bandwidths
$\Delta \omega_{\text{ls}}$ and $\Delta \omega_{\text{us}}$
centered around the lower and upper sideband respectively. (A
complete treatment based on the full Hamiltonian including all
light modes can be found in App.~\ref{DerivationME-I}.)
$\lambda_{\mathbf{k}}$ specifies the two orthogonal polarizations
of the light mode with wave vector $\mathbf{k}$. The atomic
operator
$\sigma_{I/II,i}=|\!\!\uparrow\rangle_{I/II,i}\langle\downarrow\!\!|$
refers to a particle in ensemble $I/II$ at position
$\mathbf{r}_{i}$, $\Delta\mathbf{k}=\mathbf{k}_{L}-\mathbf{k}$,
and $\mathbf{k}_{L}$ is the wavevector of the applied classical
field. AC Stark shifts have been absorbed in the detuning.
$\bar{g}(\mathbf{k})\mu$ and $\bar{g}(\mathbf{k})\nu$ are the
effective coupling strengths for the passive (beamsplitter-like)
part of the interaction and the active (squeezing) component of
the Hamiltonian respectively. More specifically,
$\bar{g}(\mathbf{k})\mu=\frac{\Omega_{\text{probe}}}{\Delta-\Omega}g_{\mathbf{k}}$
and
$\bar{g}(\mathbf{k})\nu=\frac{\Omega_{\text{probe}}}{\Delta+\Omega}g_{\mathbf{k}}$,
where $\Omega_{\text{probe}}$ is the Rabi frequency of the applied
classical field.
Here and in the following, we refer to the off-resonant
(entangling) light field as probe field. In the following sections
resonant fields are introduced which will be referred to as pump
fields.
The definition of the coupling constant for transitions between
ground and excited states $g_{\mathbf{k}}$ can be found in
App.~\ref{DerivationME-I}. The parameters
$\mu=\frac{\Delta+\Omega}{2\sqrt{\Delta\Omega}}$ and
$\nu=\frac{\Delta-\Omega}{2\sqrt{\Delta\Omega}}$ are normalized
such that $\mu^2-\nu^2=1$ \cite{Normalization}.\\
%
%
\\A Hamiltonian of type (\ref{Hamiltonian}) can be realized in many
different ways. In general, the scheme presented here can be
implemented in any system where a tunable quadratic interaction
with an active and a passive part corresponding to two sideband
modes can be realized, for example in ions or using
optopmechanical resonators.
We focus here on the creation of dissipatively driven entanglement
in atomic ensembles and continue considering the level structure
depicted in Fig. \ref{Figure2}.
If the Larmor splitting of excited states is considerably larger
than the splitting of ground states, it is for instance possible
to introduce a $\lambda/2$ plate between the two ensembles instead
of applying an external electric field \cite{HalfWavePlate}.
As discussed in Sec.~\ref{Alkali Implementation}, alkali atoms
provide another possibility to realize the desired light-matter
interaction. Due to their multi-level structure is is not even
necessary to introduce electric fields or passive optical
elements.\\
\\We remark for clarity, that the possibility illustrated in
Fig.~\ref{Figure2} implies that the effective coupling constants
(after adiabatic elimination) $\bar{g}(\mathbf{k})\mu$ and
$\bar{g}(\mathbf{k})\nu$ describing the interaction of light with
the first and the second ensemble have different signs, as the
light is blue detuned in the former and red detuned in the latter,
such that $\mu_{I}=-\mu_{II}$ and $\nu_{I}=-\nu_{II}$. This is not
the case in the implementation considered in Sec.~\ref{Alkali
Implementation}. Due to the complex levels structure, both
effective coupling constants have the same sign and therefore
$\mu_{I}=\mu_{II}$ and $\nu_{I}=\nu_{II}$. In order to describe
both alternatives in a compact way, we use a unified notation and
absorb the sign in the definition of the atomic operators
referring to the second ensembles $\sigma_{II,i}\rightarrow
\text{sgn}(\mu_{I}\mu_{II})\sigma_{II,i}$ as explained in
App.~\ref{DerivationME-II}.\\
%
%
\\It is instructive to consider Hamiltonian~(\ref{Hamiltonian}),
where excited levels have been adiabatically eliminated, because
it shows clearly that the light matter interaction depicted in
Fig.~\ref{Figure2} corresponds to a beamsplitting interaction of
the type $H\propto \int_{\Delta \omega_{ls}}d\mathbf{k}\left(A
a_\mathbf{k}^{\dag}+A^{\dag}a_\mathbf{k}\right)+\int_{\Delta
\omega_{us}}d\mathbf{k}\left(B
a_\mathbf{k}^{\dag}+B^{\dag}a_\mathbf{k}\right)$ between photons
in the upper and lower sideband with the nonlocal operators $A$
and $B$ (with additional phase factors $e^{\pm i\Delta
\mathbf{k}\mathbf{r}_i}$).
By deriving the corresponding master equation and including
thermal motion as explained in Sec.~\ref{MasterEquation3}, one can
show that this Hamiltonian yields a master equation which consists
of a desired part of type (\ref{MEideal}) with jump operators $A$
and $B$ and an additional contribution representing noise terms.
However, in the following two subsections we derive the maser
equation starting from the full Hamiltonian including excited
levels since this approach is better suited to take dipole-dipole
interactions into account.
\subsection{Effective master equation for ground states}\label{MasterEquation2}
%
%
%
In the following, we outline the derivation of the master equation
for atomic ground states $|\!\!\!\uparrow\rangle$ and
$|\!\!\!\downarrow\rangle$ and comment on the approximations used
to obtain the shown result. The full calculation can be found in
App.~\ref{DerivationME-I}.\\
\\For brevity, we use a short hand notation and abbreviate
master equations of Lindblad form $d_t\rho(t)=\kappa/2 \left(A
\rho(t)A^{\dag}-A^{\dag}A\rho(t)\right)+H.C.$ with complex decay
rate $\kappa$ and jump operator $A$ by the expression
$d_t\rho(t)=\kappa/2 A \rho(t)A^{\dag}+...\ $.\\
%
%
\\We consider the full Hamiltonian including excited levels and
undesired transitions \cite{UndesiredTransitions} without applying
the rotating wave approximation for quantum fields (see
Eq.~\ref{FullHamiltonian}). As explained in
App.~\ref{DerivationME-I}, counter-rotating terms play an
important role in the calculation the imaginary parts of the
master equation, but do not affect the real parts.
Starting from the full Hamiltonian, we obtain a master equation of
Lindblad form for the reduced atomic density matrix (see Eqs.
(\ref{ThreeParts})-(\ref{L3})). To this end, we apply the
approximation of independent rates of variation \cite{CT} and
follow the standard procedure assuming Born Markov dynamics.
The approximation of independent rates of variations is valid if
the Rabi frequency of the applied laser field
$\Omega_{\text{probe}}$ is very small compared with the
frequencies of atomic transitions. As we consider transitions in
the optical domain, this assumption is clearly legitimate. More
generally, here and in the following sections we consider
situations exhibiting two very different time scales for
variations in the system and in the bath of light modes
$\Gamma_{\text{atomic}}\tau_{\text{c}}<<1$, where
$\tau_{\text{c}}$ is the correlation time in the reservoir. For
optical frequencies this is very well justified and we can
therefore assume Born-Markov dynamics.
Moreover, we restrict ourselves to settings, where the level
splitting $\Omega$ between the states $|\!\!\uparrow\rangle$ and
$|\!\!\downarrow\rangle$ is sufficiently large, such that the
upper and lower sideband can be treated as independent baths
\cite{IndependentBaths} (compare App.~\ref{DerivationME-I}).
Finally, we assume that the condition $k_{L}\gg R/L^2$ is
fulfilled. $k_{L}$ is the wave vector of the applied laser field.
Since we consider frequencies in the optical domain, $k_{L}$ is on
the order of $10 ^7\text{m}^{-1}$. $L$ is the spatial extent of
the atomic ensembles, which we consider to be on the order of cm,
while the distance between the two ensembles $R$ is about one
meter.\\
%
%
%
\\As next step, excited states are adiabatically eliminated under
the condition $|\Delta|\gg \Gamma_{\text{atomic}}, \
\delta_{\text{Doppler}}$. This leads to an effective master
equation for atomic ground states. Using the abbreviated notation
introduced in the beginning of this subsection,
\begin{eqnarray}
d_{t}\rho(t)\!\!\!&=&\!\!\!\frac{1}{2}\sum_{i,j=1}^N e^{-i
\mathbf{k}_{L}\left(\mathbf{r}_{j}-\mathbf{r}_i\right)}J_{ij}\left(
A_i\rho(t) A_{j}^{\dag} + B_i\rho(t)
B_{j}^{\dag}\right)\nonumber\\
\!\!\!&+&\!\!\!\frac{1}{2}\sum_{i,j=1}^N e^{-i
\mathbf{k}_{L}\left(\mathbf{r}_{j}-\mathbf{r}_i\right)}\check{J}_{ij}\left(C_i\rho(t)
C_{j}^{\dag}+D_i\rho(t) D_{j}^{\dag}\right)\nonumber\\
\!\!\!&+&\!\!\!... \label{MEafterAdiabaticElimination}
\end{eqnarray}
where $J_{ij}$ and $\check{J}_{ij}$ are complex decay rates which
are discussed below and $A=\frac{1}{\sqrt{N}}\sum_{i=1}^{N}A_{i}$.
The operators $B$, $C$ and $D$ are analogously defined as sums.
$A_i$, $B_i$, $C_i$ and $D_i$ are given by
\begin{eqnarray}\label{Operators}
A_i&=&\mu \ \sigma_{I,i}+\nu\ \sigma^{\dag}_{II,i},\\
B_i&=&\mu\  \sigma_{II,i}+\nu\ \sigma^{\dag}_{I,i},\nonumber\\
C_i&=&\mu\  \sigma_{\downarrow\downarrow,I,i}+\nu\ \sigma_{\uparrow\uparrow,II,i},\nonumber\\
D_i&=&\mu\ \sigma_{\downarrow\downarrow,II,i}+\nu\
\sigma_{\uparrow\uparrow,I,i},\nonumber
\end{eqnarray}
where the abbreviations
$\sigma_{\uparrow\uparrow,I/II,i}=|\uparrow\rangle_{I/II,i}\langle
\uparrow|$ and
$\sigma_{\downarrow\downarrow,I/II,i}=|\downarrow\rangle_{I/II,i}\langle
\downarrow|$ are used. Terms involving the operators $A$ and $B$
represent desired transitions involving a spin flip
$|\!\!\uparrow\rangle \rightarrow |\!\!\downarrow\rangle$ or
$|\!\!\downarrow\rangle \rightarrow |\!\!\uparrow\rangle$ as shown
in Fig.~\ref{Figure2}. Terms involving the operators $C$ or $D$
represent undesired transitions which lead to dephasing
\cite{UndesiredTransitions}.
Desired and undesired transitions are associated with different
decay rates $J_{ij}$ and $\check{J}_{ij}$ respectively. In the
four level model considered here, $\check{J}_{ij}=2J_{ij}$, due to
the ratio of Clebsch Gordan coefficients $c^2_{\Delta_m=\pm
1}/c^2_{\Delta_m=0}=2$.
As introduced above, $J_{ij}=\gamma(\mathbf{r}_{ij})+i
g(\mathbf{r}_{ij})$ is a complex decay rate with real part
$\gamma(\mathbf{r}_{ij})=\gamma(\mathbf{r}_{ji})$ and imaginary
part $g(\mathbf{r}_{ij})=g(\mathbf{r}_{ji})$. Imaginary single
particle terms represent energy shifts (single atom Lamb shift)
and are absorbed in the definition of detunings. Therefore we
consider in the following only imaginary terms
$g(\mathbf{r}_{ij})$ with $i\neq j$ and use renormalized atomic
energies and the resulting effective detunings.
The real and imaginary part of $J_{ij}$ are given by
\cite{Lehmberg}
\begin{eqnarray}
\gamma(\mathbf{r}_{ij})\!\!\!&=&\!\!\!\frac{3}{2}\Gamma
\!\left(1\!-\!\left(\hat{\mathbf{p}}\cdot \hat{\mathbf{r}}_{ij}
\right)^2\right)\!\frac{\sin(k_{\text{L}} r_{ij})}{k_{\text{L}}
r_{ij}}\label{RealPart}\\
\!\!\!&+&\!\!\!\frac{3}{2}\Gamma\!\left(1-3\left(\hat{\mathbf{p}}\cdot
\hat{\mathbf{r}}_{ij}
\right)^2\right)\!\!\left(\frac{\cos(k_{\text{L}}
r_{ij})}{(k_{\text{L}} r_{ij})^2}\!-\!\frac{\sin(k_{\text{L}}
r_{ij})}{(k_{\text{L}} r_{ij})^3}\right)\!,\nonumber\\
g(\mathbf{r}_{ij})\!\!\!&=&\!\!\!-\frac{3}{2}\Gamma
\!\left(1-\left(\hat{\mathbf{p}}\cdot \hat{\mathbf{r}}_{ij}
\right)^2\right)\frac{\cos(k_{\text{L}} r_{ij})}{k_{\text{L}}
r_{ij}}\label{ImaginaryPart}\\
\!\!\!&+&\!\!\!\frac{3}{2}\Gamma\!\left(1-3\left(\hat{\mathbf{p}}\cdot
\hat{\mathbf{r}}_{ij}
\right)^2\right)\!\!\left(\frac{\sin(k_{\text{L}}
r_{ij})}{(k_{\text{L}} r_{ij})^2}\!+\!\frac{\cos(k_{\text{L}}
r_{ij})}{(k_{\text{L}} r_{ij})^3}\right)\!,\nonumber
\end{eqnarray}
where $\hat{\mathbf{p}}$ is the unit vector of the dipole matrix
element $\mathbf{p}=\langle e_{\uparrow}|e\
\hat{\textbf{x}}|\uparrow\rangle$, which we assume to be real.
$\hat{\textbf{r}}_{ij}$ is the unit vector of the interatomic
distance $\mathbf{r}_{ij}=\mathbf{r}_i-\mathbf{r}_j$ and
$r_{ij}=r_{ji}$ is the length of the vector $\mathbf{r}_{ij}$.
$\Gamma$ is the effective decay rate of a single isolated atom.
%
%
%
\subsection{Master equation including thermal
motion and noise processes}\label{MasterEquation3}
%
%
%
In this subsection, atomic motion is taken into account
\cite{DiegoI,DiegoII,DuanCiracZoller02}. As is shown below,
thermal motion gives rise to noise terms which are small compared
the desired
contributions for samples with high optical depth.\\
%
%
\\Atoms are statistically distributed. The dynamics of the whole
system is thereby governed by two different time scales, the
characteristic time of radiative emission $1/
\Gamma_{\text{atomic}}$ and the characteristic time of atomic
redistribution $\frac{L}{v}$, where $L$ is the length of a cubic
ensemble and $v$ is the average velocity of particles. In the
limit, where the time scale of atomic motion is fast compared to
the time scale of radiative decay $\Gamma_{\text{atomic}}
\frac{L}{v} \ll 1$ one can describe the emission independently of
the evolution of atomic positions which enters the master equation
in the form of averaged coefficients, where the average in time
corresponds to an average in space \cite{Oxana}. Atomic positions
can be treated as independent random variables and for simplicity,
we choose a Gaussian probability distribution of width $L$,
$P(r)=\frac{1}{\pi^{3/2}L^3}e^{-\frac{r^2}{L^2}}$
\cite{DuanCiracZoller02}.
As shown in App.~\ref{DerivationME}, we find that imaginary parts
of the averaged decay rates can be neglected. The averaged master
equation is given by
\begin{eqnarray}
d_{t}\rho(t)&=&\frac{1}{2}\sum_{i,j=1}^N
\Gamma_{ij}\left(A_i\rho(t) A_{j}^{\dag}
+B_i\rho(t)B_{j}^{\dag}\right) \nonumber\\
&+&\frac{1}{2}\sum_{i,j=1}^N \check{\Gamma}_{ij}\left(C_i\rho(t)
C_{j}^{\dag}
+D_i\rho(t)D_{j}^{\dag}\right)\nonumber\\
&+&...\ .\label{MasterEquationAferAveraging}
\end{eqnarray}
with $\check{\Gamma}_{ij}=2\Gamma_{ij}$ for the basic model
discussed here.
For $k_{L} L\gg 1$ and $i\neq j$,
$\Gamma_{ij}=\Gamma\frac{3}{4\left(k_L L\right)^2}$.
$d=N\frac{\Gamma}{\Gamma_{ij}}=\frac{3N}{4 \left(k_L L\right)^2}$
is the resonant optical depth of one atomic ensemble.
Using this definition and $1/\left(k_L L\right)^2\ll 1$, one
obtains
\begin{eqnarray}\label{ME2}
d_{t}\rho(t)&=&d\frac{\Gamma}{2} A \rho(t)
A^{\dag} + d\frac{\Gamma}{2}  B \rho(t) B^{\dag}\\
&+&\mu^2\frac{\Gamma}{2}\sum_{i=1}^{N}\left(
\sigma_{I,i}\rho(t)\sigma_{I,i}^{\dag}+
\sigma_{II,i}\rho(t)\sigma_{II,i}^{\dag}\right)\nonumber\\
&+&\nu^2\frac{\Gamma}{2}\sum_{i=1}^{N}\left(
\sigma_{I,i}^{\dag}\rho(t)\sigma_{I,i}+
\sigma_{II,i}^{\dag}\rho(t)\sigma_{II,i}\right)\nonumber\\
&+&d\frac{\check{\Gamma}}{2} C \rho(t) C^{\dag} +
d\frac{\check{\Gamma}}{2} D \rho(t)D^{\dag}
+\frac{\check{\Gamma}}{2}\left(\mu^2+\nu^2\right)\nonumber\\
&&\sum_{i=1}^{N}\!\!\left(\sigma_{\downarrow\downarrow,I,i}\rho(t)\sigma_{\downarrow\downarrow,I,i}
+\sigma_{\downarrow\downarrow,II,i}\rho(t)\sigma_{\downarrow\downarrow,II,i}\right)\nonumber\\
&+&...\ .\nonumber
\end{eqnarray}
The first three lines correspond to the first sum in
Eq.~(\ref{MasterEquationAferAveraging}).
The entangling terms in the first line are enhanced by a factor
$d$. For sufficiently optically thick samples, additional noise
terms in the second and third line, which reflect thermal motion,
are small compared to the desired contributions.
The last two lines correspond to the second sum in
Eq.~(\ref{MasterEquationAferAveraging}), where
$|\!\!\uparrow\rangle\langle
\uparrow\!\!|+|\!\!\downarrow\rangle\langle
\downarrow\!\!|=1\!\text{I}$ was used. The first two terms
$d(\check{\Gamma}/2) C \rho(t) C^{\dag} + d(\check{\Gamma}/2) D
\rho(t)D^{\dag}$ are collective dephasing terms. They do not have
an effect on the entanglement generated (see App.~\ref{FullME})
and can therefore be omitted in the following.\\
\\In the following sections, the effect of pump fields is
considered. Resonant pump fields cause incoherent cooling (and
heating) processes, which can be taken into account by adding
cooling (and heating) terms which correspond to a transfer of
atoms from level $|\!\!\downarrow\rangle$ to level
$|\!\!\uparrow\rangle$ (and back). Finally, we include additonal
processes, which do not lead to spin flips but cause dephasing,
such as fluctuating magnetic fields. The full master equation is
given by
\begin{eqnarray}\label{ME3}
d_{t}\rho(t)\!\!&=&\!\! \!d\frac{\Gamma}{2} A \rho(t)
A^{\dag} + d\frac{\Gamma}{2}  B \rho(t) B^{\dag}\\
\!\!&+&\!\!\!\frac{\Gamma_{\rm{cool}}}{2}\sum_{i=1}^{N}\!\!\left(
\sigma_{I,i}\rho(t)\sigma_{I,i}^{\dag}\!+\!
\sigma_{II,i}\rho(t)\sigma_{II,i}^{\dag}\right)\nonumber\\
\!\!&+&\!\!\!\frac{\Gamma_{\rm{heat}}}{2}\sum_{i=1}^{N}\!\!\left(
\sigma_{I,i}^{\dag}\rho(t)\sigma_{I,i}\!+\!
\sigma_{II,i}^{\dag}\rho(t)\sigma_{II,i}\right)\nonumber\\
\!\!&+&\!\!\!\frac{\Gamma_{\!\!\rm{d}}}{2}\sum_{i=1}^{N}\left(\sigma_{\downarrow\downarrow,I,i}\rho(t)\sigma_{\downarrow\downarrow,I,i}
+\sigma_{\downarrow\downarrow,II,i}\rho(t)\sigma_{\downarrow\downarrow,II,i}\right)\nonumber\\
&+&...\ .\nonumber
\end{eqnarray}
Note that the last three lines represent single particle
processes. They do not feature a collective enhancement factor $d$
as the entangling terms in the first line. The noise terms
proportional to $\Gamma\mu^2$ and $\Gamma\nu^2$ in the second and
third line in expression (\ref{ME2}) have been absorbed in lines
two and three of Eq.~(\ref{ME3}). Hence, $\Gamma_{\rm{cool}}$
($\Gamma_{\rm{heat}}$) is the total single-particle cooling
(heating) rate. Noise terms proportional to $\Gamma
\left(\mu^2+\nu^2\right)$ in expression (\ref{ME2}) have been
absorbed in the last line, such that $\Gamma_{\!\!\rm{d}}$ is the
total dephasing rate. More details concerning the derivation of
the full master equation (\ref{ME3}) can be found in
App.~\ref{FullME}.
\subsection{Creation of entanglement}\label{CreationOfEntanglement}
\begin{figure}[pbt]
\begin{center}
\includegraphics[width=7cm]{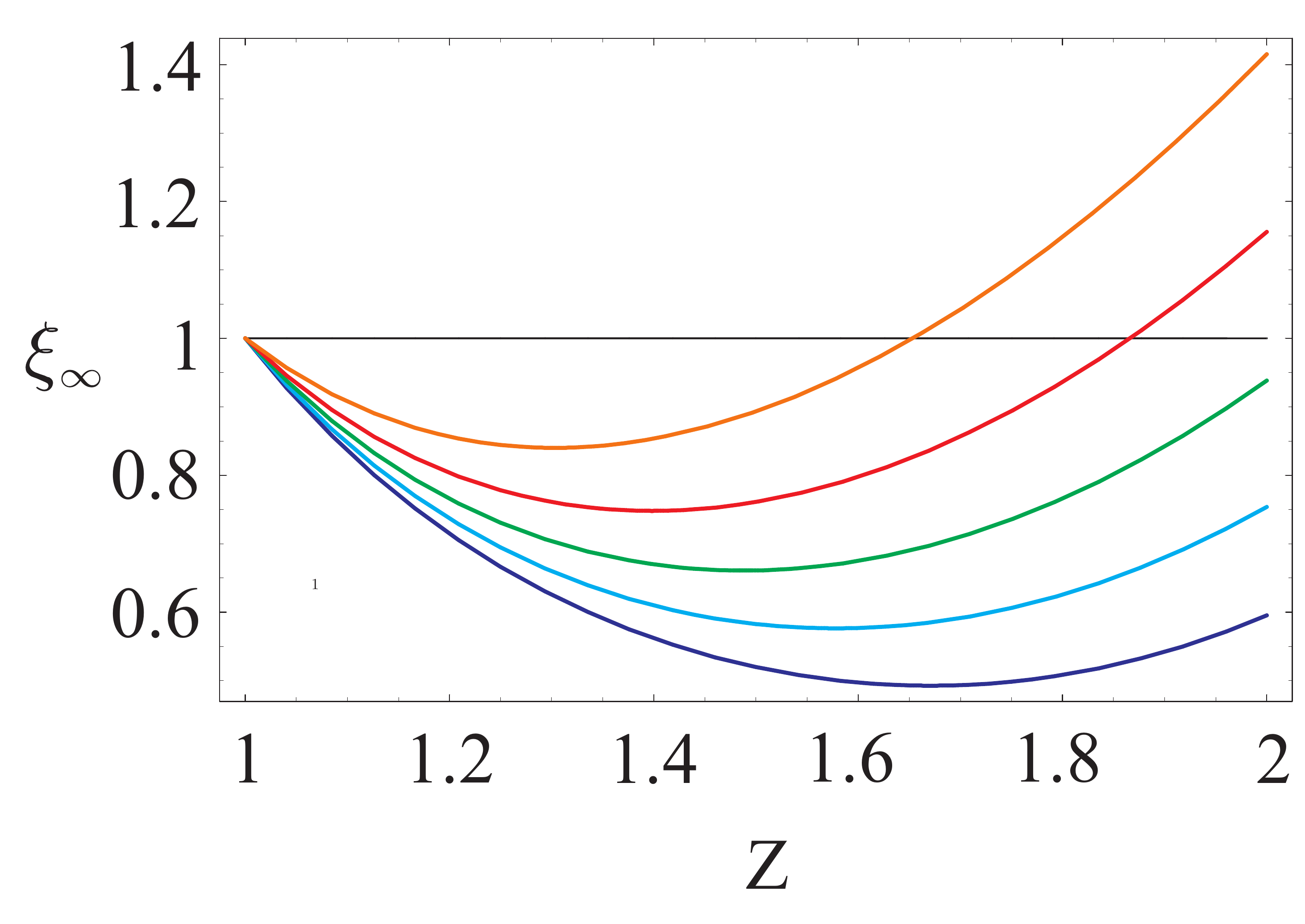} \caption{(Color online) Steady state
entanglement $\xi_{\infty}$ in a two-level system versus
$Z=\left(|\mu|-|\nu|\right)^{-1}$ for optical depth $d=30$ per
ensemble. The horizontal black line indicates the separable limit.
(For separable states $\xi\geq 1$, the smaller $\xi$, the higher
the amount of entanglement) The lowest line (violet) depicts
$\xi_{\infty}$ for purely radiative dephasing
$\Gamma_{\text{d}}^{\text{add}}=0$. The next curves show in
ascending order $\Gamma_{d}^{\text{add}}=2\Gamma$ (blue),
$\Gamma_{\text{d}}^{\text{add}}=5\Gamma $ (green),
$\Gamma_{\text{d}}^{\text{add}}=10\Gamma $ (red) and
$\Gamma_{\text{d}}^{\text{add}}=20\Gamma $ (orange), where
$\Gamma$ is the single particle decay rate.}\label{TwoLevelPlot}
\end{center}
\end{figure}
\begin{figure}[pbt]
\begin{center}
\includegraphics[width=7.5cm]{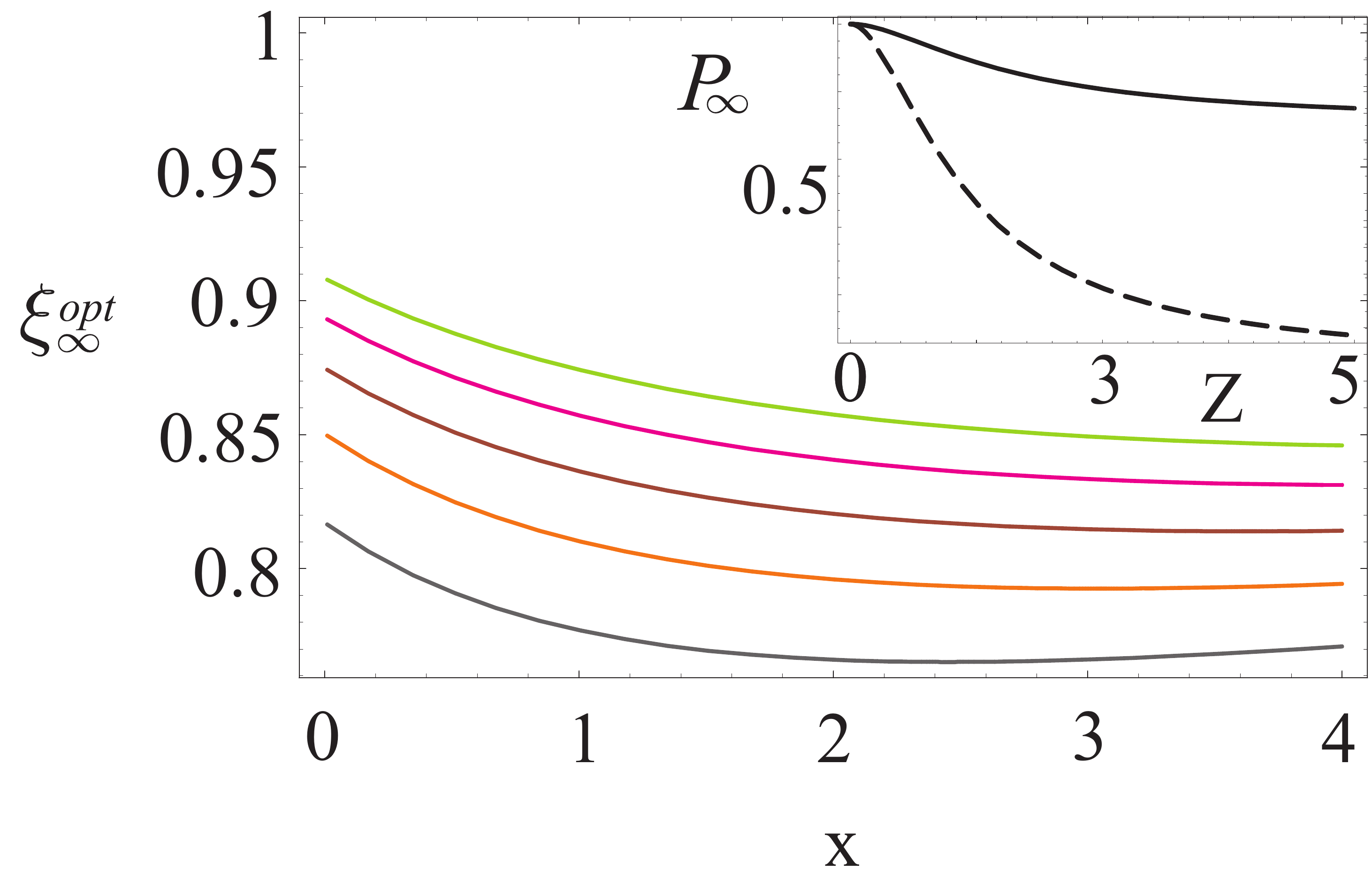} \caption{(Color online) Steady state
entanglement $\xi_{\infty}^{\text{opt}}$ for optimal squeezing
parameter $Z$ versus pump parameter $x$. (Two-level model,
$d=30$). The curves correspond in ascending order to
$\Gamma_{\text{d}}^{\text{add}}=15$(grey),
$\Gamma_{\text{d}}^{\text{add}}=20\Gamma$ (orange),
$\Gamma_{\text{d}}^{\text{add}}=25\Gamma $ (brown),
$\Gamma_{\text{d}}^{\text{add}}=30\Gamma $ (pink) and
$\Gamma_{\text{d}}^{\text{add}}=35\Gamma $ (light green). The
inset shows the steady state polarization $P_{2,\infty}$ versus
$Z$ in the absence of pump fields $x=0$ (dashed line) and for
$x=5$ (solid line).}\label{PumpPlot}
\end{center}
\end{figure}

In this subsection, we determine how much entanglement can be
generated by the proposed scheme in the presence of noise
processes for a given optical depth $d$ and given parameters $\mu$
and $\nu$. Details of the calculation can be found in
App.~\ref{Appendix1}. For simplicity, we assume identical
conditions for both ensembles. The amount of entanglement produced
is measured by means of the quantity $\xi$ defined in
Eq.~(\ref{xi}). Hence, the time evolution of $\Sigma_J=\rm {var}
(J_{y,I} + J_{y,II})+\rm {var} (J_{z,I} - J_{z,II})$ as well as
the evolution of the mean value of the longitudinal spin $|\langle
J_{x,I}\rangle| = |\langle J_{x,II}\rangle|$ need to be
calculated. We consider the limit $N\gg 1$ and start by
computing the former.\\
\\ $\Sigma_{J}$ decays according to
\begin{eqnarray*}
d_t\Sigma_J&=&-\left(\tilde{\Gamma}+d\Gamma
P_2(t)\right)\Sigma_J+N d\Gamma
P_2(t)^2\left(|\mu|-|\nu|\right)^2,
\end{eqnarray*}
where
$\tilde{\Gamma}=\Gamma_{\rm{cool}}+\Gamma_{\rm{heat}}+\Gamma_{\text{d}}$
and $P_2(t)=\frac{2}{N}\langle J_x\rangle$. The evolution of the
mean value of the longitudinal spin is given by
\begin{eqnarray*}
d_{t}\langle
J_x\rangle=-\frac{1}{2}\left(\Gamma_{\rm{heat}}+\Gamma_{\rm{cool}}\right)\langle
J_x\rangle_t+ \frac{N}{2}
\left(\Gamma_{\rm{cool}}-\Gamma_{\rm{heat}}\right).
\end{eqnarray*}
There are two distinct time scales. For atomic ensembles with high
optical depth, the evolution of the transverse spin components is
collectively enhanced and therefore fast compared to the decay of
$\langle J_x\rangle$ which is due to single particle processes.
In the limit where the entangled quantum state follows the
changing atomic polarization adiabatically, the time evolution of
$\xi(t)$ is given by
\begin{eqnarray}
 \xi(t)\!\!&=&\!\!\frac{1}{P_2(t)}e^{-\left(\tilde{\Gamma}+d\Gamma
P_2(t)\right)t}\\
\!\!&+&\!\!\frac{1}{P_2(t)}\frac{\tilde{\Gamma}+d\Gamma
P_2(t)^2(|\mu|-|\nu|)^2}{\tilde{\Gamma}+d\Gamma
P_2(t)}\left(1-e^{-\left(\tilde{\Gamma}+d\Gamma
P_2(t)\right)t}\right)\!.\nonumber\label{TimeEvolution}
\end{eqnarray}
In the steady state
\begin{eqnarray}\label{Main result}
 \xi_{\infty}&=&
 \frac{1}{P_{2,\infty}}\frac{\tilde{\Gamma}
 +d\Gamma
 P_{2,\infty}^2\left(|\mu|-|\nu|\right)^2}{\tilde{\Gamma}+d\Gamma
 P_{2,\infty}},\\
 P_{2,\infty}&=&\frac{\Gamma_{\rm{cool}}-\Gamma_{\rm{heat}}}{\Gamma_{\rm{cool}}+\Gamma_{\rm{heat}}}.\nonumber
\end{eqnarray}
This result shows that for high optical depth, the system reaches
an entangled steady state. Under the dissipative dynamics
considered here, entanglement persists for arbitrarily long times.
In the absence of noise, $\tilde{\Gamma}=0$ and Eq.~(\ref{Main
result}) reduces to $
\xi_{\infty}=\left(|\mu|-|\nu|\right)^2$.\\
\\Fig.~\ref{TwoLevelPlot} shows the attainable amount of
entanglement in the steady state $\xi_{\infty}$ for moderate
optical depth $d=30$ versus $Z=\left(|\mu|-|\nu|\right)^{-1}$ if
only probe fields are applied. In this case
$\Gamma^{\text{probe}}_{\text{cool}}=\mu^2\Gamma$ and
$\Gamma^{\text{probe}}_{\text{heat}}=\nu^2\Gamma$. The dephasing
rate
$\Gamma_{\text{d}}=\Gamma_{\text{d}}^{\text{rad}}+\Gamma_{d}^{\text{add}}$
consists of a radiative part
$\Gamma_{\text{d}}^{\text{rad,probe}}=2\left(\mu^2+\nu^2\right)\Gamma$
\cite{2Deph}, which is due to light-induced transitions
$|\!\!\uparrow\rangle\rightarrow|\!\!\uparrow\rangle$ and
$|\!\!\downarrow\rangle\rightarrow|\!\!\downarrow\rangle$, and an
additional term $\Gamma_{\text{d}}^{\text{add}}$ which summarizes
all non-radiative sources of dephasing such as fluctuating
magnetic fields. This additional component can take values up to
$\Gamma_{\text{d}}^{\text{add}}=20\Gamma$ while still allowing for
a reduction of $\xi_{\infty}$ by $15\%$.
For large values of $\Gamma_{d}^{\text{add}}$, the limiting
mechanism is the decrease in polarization for high squeezing
parameters and can be counteracted by applying resonant
$\sigma_{+}$ and $\sigma_{-}$ polarized pump fields to the first
and second ensemble respectively, which drive the transition
$|\!\!\downarrow\rangle\rightarrow|e_{\uparrow}\rangle$. In this
case, the cooling rate can be roughly estimated as
$\Gamma_{\text{cool}}=\left(1+x\right)\Gamma \mu^2$ \cite{Rates}.
The pump parameter $x$ is given by
$x=\frac{\Omega^2_{\text{pump}}}{\gamma_{\text{LW}}^2}\frac{\left(\Delta-\Omega\right)^2}{\Omega_{\text{probe}}^2}
k$, where $\Omega_{\text{pump}}$ is the Rabi frequency of the pump
field and $\gamma_{\text{LW}}$ is the natural line width of
excited levels. The correction factor $k$ takes Doppler broadening
due to thermal motion into account \cite{Doppler}.
In the presence of pump fields, radiative dephasing is enhanced
$\Gamma_{\text{d}}^{\text{rad}}=2\left(\left(1+x\right)\mu^2+\nu^2\right)\Gamma$.
The heating rate is unaffected. Fig.~\ref{PumpPlot} shows the
maximal attainable amount of entanglement
$\xi_{\infty}^{\text{\text{opt}}}$ (entanglement for optimal
squeezing parameter $Z$), for $d=30$ versus pump parameter $x$.
For $x=5$, additional dephasing up to
$\Gamma_{\text{d}}^{\text{add}}=37\Gamma$ can be tolerated while
still allowing for a reduction of $\xi_{\infty}^{\text{opt}}$ by
$15\%$. Remarkably, the application of external pump fields, which
amounts to adding extra noise to the system, is advantageous in
this case.
%
%
%
\section{Implementation in multi-level systems}\label{Alkali Implementation}
In Sec.~\ref{TwoLevelModel}, the theoretical framework for
creating steady state entanglement between two atomic ensembles at
room temperature is presented in detail for two-level systems.
This section complements the main results derived in
Sec.~\ref{TwoLevelModel} by considering the implementation in
multi-level systems. In the following we investigate the
possibility of transferring the concepts developed for two-level
systems to atoms with multi-level ground states by means of a
general simplified model and analyze the conditions for obtaining
entanglement in a steady state
qualitatively.\\
\\As specific example, we study the implementation of the proposed
scheme in ensembles of alkali atoms, where the two-level system is
encoded in a multi-level ground state manifold. Due to the richer
internal structure, no external electric fields or optical
elements need to be employed in contrast to the setup discussed in
the previous section. As explained below, suitable values
$\mu_{I}=\mu_{II}$ and $\nu_{I}=\nu_{II}$ are realized naturally.
In principle, it is possible to include all magnetic sublevels and
all possible transitions of a particular alkali atom in the
following consideration. However, rather than aiming for a
complete description which takes the entire level structure of a
specific atom into account, the general model used here is
primarily intended to describe the underlaying physics. In
Sec.~\ref{MultilevelDynamics}, we show how additional dynamics in
a multi-level system can be taken into account by means of this
simplified model which allows us to describe the physical effects
with a small set of parameters while capturing all relevant
features. In Sec.~\ref{MultilevelCalculation} we estimate the
attainable entanglement.
\subsection{Including multi-level dynamics}\label{MultilevelDynamics}
In the following we consider encoding of a two-level subsystem in
a multi-level ground state. For example, the ground state of
alkali atoms with nuclear spin $I$ is split in two manifolds with
total angular momentum $F=I+1/2$ and $F'=I-1/2$ respectively. The
relevant two-level subsystem can be encoded in the two outermost
states of the $F=I+1/2$ ground state manifold $|\!\!\uparrow
\rangle \equiv |F,\pm F\rangle$ and $|\!\!\downarrow \rangle
\equiv |F,\pm(F-1)\rangle$ in the first/second ensemble.
\\In general, the maximum attainable amount of entanglement
$\xi^{\text{ideal}}=\left(|\mu|-|\nu|\right)^2$ is determined by
the different rates $\mu^2 \Gamma$ and $\nu^2 \Gamma$ at which
probe-field induced transitions $| \!\!\downarrow\rangle
\rightarrow |\!\!\uparrow \rangle$ and $| \!\!\uparrow\rangle
\rightarrow |\!\!\downarrow \rangle$ occur. The values of the
parameters $\mu$ and $\nu$ depend on the multi-level structure of
the excited states as well as on polarization and detuning of the
applied laser field.
\begin{figure}[pbt]
\begin{center}
\includegraphics[width=8cm]{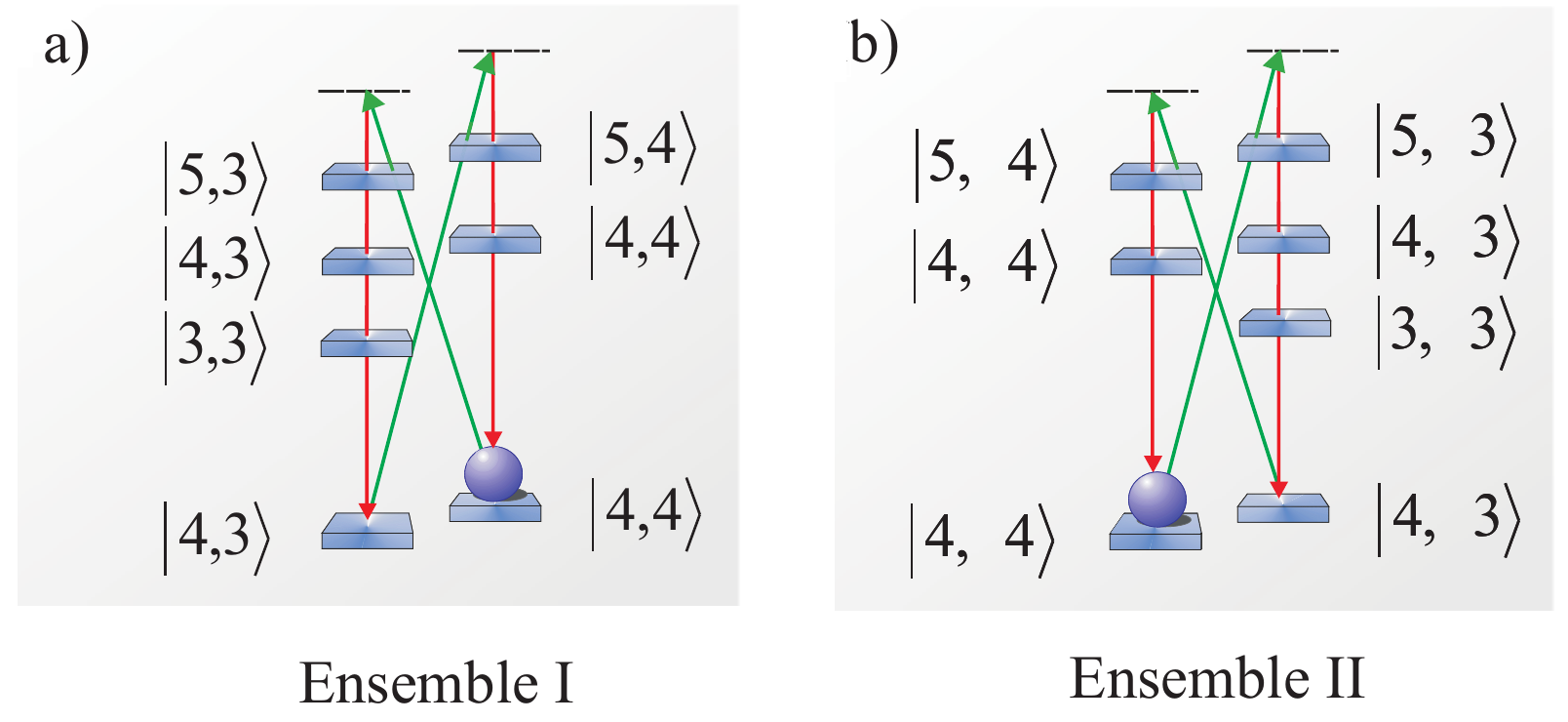}
\caption{(Color online) Implementation of the proposed scheme in
$^{133}\text{Cs}$ ensembles. The atomic samples are placed in
$\hat{\mathbf{x}}$ polarized magnetic fields and interact with a
$\hat{\mathbf{y}}$ polarized probe field propagating along
$\hat{\mathbf{z}}$ as shown in Fig.~\ref{Figure1}. The relevant
two-level subsystem is encoded in the $6S_{1/2}$ ground state,
$|\!\!\uparrow\rangle\equiv|4,\pm 4\rangle$ and
$|\!\!\downarrow\rangle\equiv|4,\pm 3\rangle$ in the first/second
ensemble and are coupled to the excited states in $6P_{3/2}$. Only
desired transitions are shown. Level splittings are not
represented true-to-scale. (In a magnetic field of $1$ Gauss, the
Zeeman shift of magnetic sublevels is about $10^5$Hz while the
hyperfine, -and fine splitting is on the order of $10^8$ Hz and
$10^{14}$ respectively.)}\label{Cesiumlevels}
\end{center}
\end{figure}
This can be illustrated by considering the off-resonant probing of
$^{133}\text{Cs}$ atoms on the $\text{D}_2$ line using the setup
shown in Fig.~\ref{Figure1}. $\hat{\mathbf{y}}$-polarized probe
light which propagates along $\hat{\mathbf{z}}$, interacts
successively with two Cs ensembles in $\hat{\mathbf{x}}$-oriented
magnetic fields. We assume weak magnetic fields ($B\approx
1$Gauss), such that the Larmor splitting $\Omega\approx 300$kHz is
much smaller than the fine splitting of excited states. The first
ensemble is strongly spin polarized along the orientation of the
magnetic field, while the second ensemble is polarized
antiparallel. As shown in Fig.~\ref{Cesiumlevels}, the passive
interaction, which transfers atoms from $|\!\!\downarrow\rangle$
to $|\!\!\uparrow\rangle$ involves the upper levels with $F=4,5$,
whereas the active part of the light-matter interaction
$|\!\uparrow\rangle\rightarrow |\!\downarrow\rangle$ involves the
manifolds with $F=3,4,5$. Taking the different Clebsch Gordan
coefficients into account, one obtains
$Z=\left(\mu-\nu\right)^{-1}=2.3$ for blue detuning
$\Delta=700\text{MHz}$ with respect to the $F=5$ manifold of
$6P_{3/2}$. (Both parameters, $\mu$ and $\nu$, are positive.)
Alternatively, $\hat{\mathbf{x}}$ polarized probe light can be
used in combination with red detuning as shown in
Fig.~\ref{Alkalilevels}. In this case $Z=2.4$ for
$\Delta=-700\text{MHz}$. Both variants are possible; in the
following we will consider $\hat{\mathbf{x}}$ polarized probe fields.\\
\\More generally, the multi-level structure of excited states affects only the
value of $\xi^{\text{ideal}}$. The multi-level character of the
ground state leads to additional dynamics that needs to be taken
into account. Firstly, atoms can leave the relevant two-level
subsystem. While matter and light interact, atoms redistribute or
are lost to other groundstate manifolds. Therefore, the atomic
population in the two-level subsystem is continually reduced. This
is accounted for by introducing a time dependent population
$N_2(t)$ and including the effect accordingly in the corresponding
polarization $P_2(t)$. The subscript "2" emphasizes that these
quantities are defined with respect to the two-level subsystem
$\{|\!\!\uparrow\rangle,|\!\!\downarrow\rangle\}$. In order to
calculate $N_2(t)$ and $P_2(t)$ in
Sec.~\ref{MultilevelCalculation}, we introduce now a general model
which allows one to analyze the realization of the proposed scheme
in atoms with multi-level ground states. A high degree of
population and polarization with respect to the two-level
subsystem is required in the process of generating long-lived
entanglement. Therefore $\sigma_{\pm}$ polarized pump and repump
fields have to be applied. These additional fields induce
transitions with $\Delta m_{F}=+1$ in the first ensemble and
transitions with $\Delta m_{F}=-1$ in the second one. For alkali
ensembles, pump fields drive transitions within the manifold
$F=I+1/2$ while repump fields transfer atoms in $F'=I-1/2$ back to
$F=I+1/2$. In the desired case of high polarization with respect
to the two outermost states, the atomic population in sublevels
with $F=4$, $\pm m_F<3$ in the first/second ensemble can be
neglected. In this regime it is sufficient to restrict the
description to three states, $|\!\!\uparrow\rangle$,
$|\!\!\downarrow\rangle $ and $|h\rangle \equiv |F',F'\rangle$ for
the first and $|h\rangle \equiv
|F',-F'\rangle$ for the second ensemble, as shown in Fig.~\ref{Alkalilevels}.\\
\\Finally, one has to distinguish between spin operators which
refer to the relevant two-level subsystem and experimentally
measurable quantities which are defined with respect to $F=I+1/2$.
For clarity, operators referring to the full multi-level structure
are labelled with the subscript "exp". The longitudinal spin of
each ensemble is given by
\begin{eqnarray}
J_{x,\rm{exp}}\!\!&=&\!\!\sum_{i=1}^N\sum_{m=-F}^F\!\! m
|m\rangle_i\langle m|\\
&\approx& \sum_{i=1}^{N}\left(F|\!\!\uparrow\rangle_{i}\langle
\uparrow|+(F-1)|\!\!\downarrow\rangle_{i}\langle \downarrow|\right)\nonumber\\
\!\!&=&\!\! J_{x,2}+\frac{2F-1}{2}N_2(t),\label{Jx}
\end{eqnarray}
where $N_2=\sum_{i=1}^N\left(|\!\!\uparrow\rangle_{i}\langle
\uparrow|+|\!\!\downarrow\rangle_{i}\langle \downarrow|\right)$.
For transverse spin components,
\begin{eqnarray}
 J_{y,\rm{exp}}\!\!&=&\!\!\frac{1}{2}\sum_{i=1}^{N}\sum_{m=-F}^F\!\!\sqrt{F(F+1)-m(m+1)}(|m+1\rangle_i\langle m|\nonumber\\
 \!\!&+&\!\!|m\rangle_i\langle
 m+1|)\approx\sqrt{2F}J_{y,2},\label{Jy}
\end{eqnarray}
such that
\begin{eqnarray}
\Sigma_{J,\text{exp}}=2
F\Sigma_{J,2}+2(2F-1)N_{\downarrow},\label{TransVar}
\end{eqnarray}
with $N_{\downarrow}=\sum_{i=1}^{N}|\!\!\downarrow\rangle_i\langle
\downarrow\!\!|$.
\begin{figure}[pbt]
\begin{center}
\includegraphics[width=8cm]{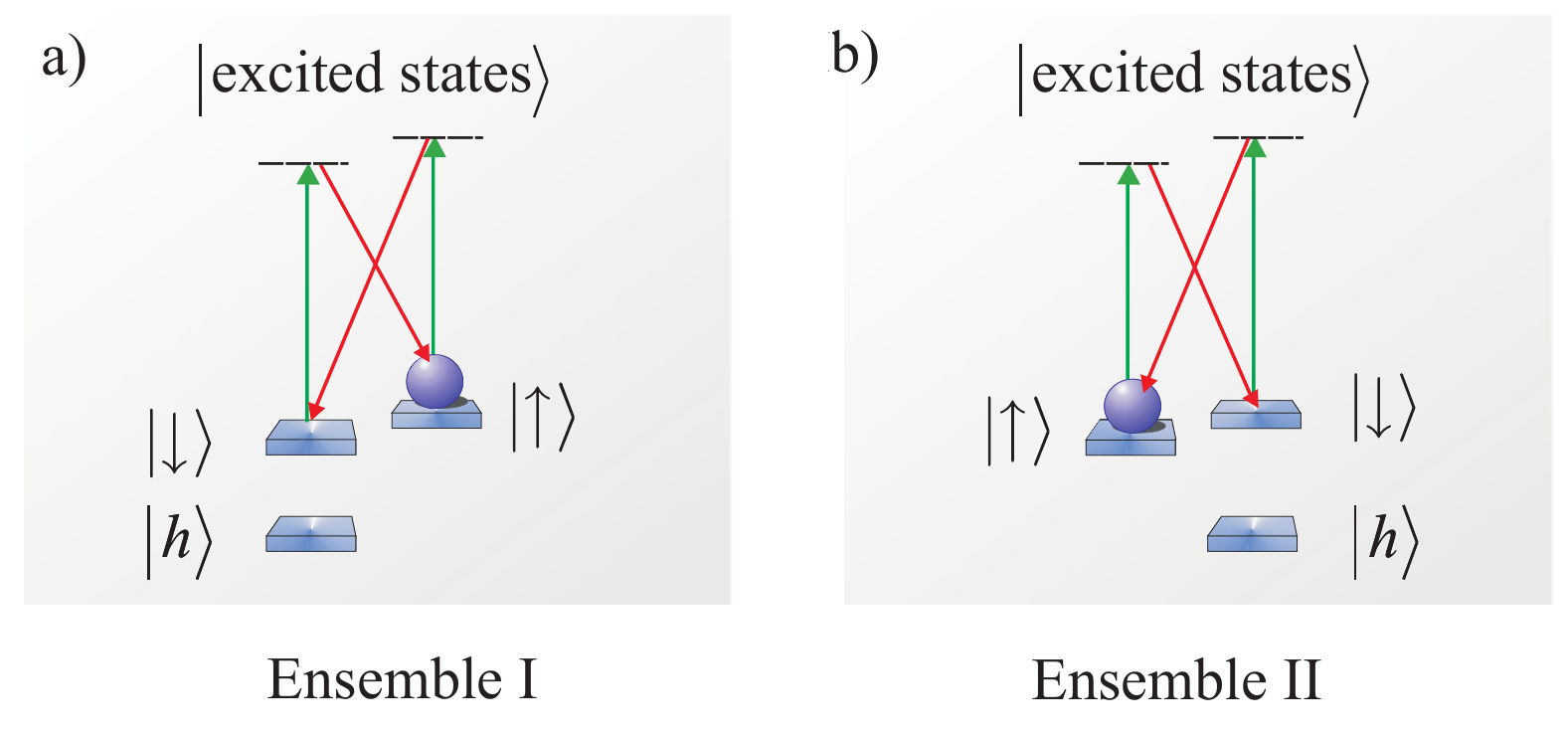}
\caption{(Color online) General scheme for the realization of
long-lived entanglement between alkali-ensembles. The polarization
of the probe field is assumed to be parallel to the applied
magnetic field. The ground state $S_{1/2}$ consists of two
manifolds with total spin $F=I+1/2$ and $F'=I-1/2$, where $I$ is
the nuclear spin. The states $|\!\!\uparrow\rangle$ and
$|\!\!\downarrow\rangle$ are encoded in the outermost levels of
the $F=I+1/2$ ground state manifold. We identify
$|\!\!\uparrow\rangle\equiv |F,F\rangle$ and
$|\!\!\downarrow\rangle\equiv |F,F-1\rangle$ for the first atomic
ensemble and $|\!\!\uparrow\rangle\equiv |F,-F\rangle$ and
$|\!\!\downarrow\rangle\equiv |F,-F+1\rangle$ for the second one.
Only desired transitions are shown. In the presence of strong pump
fields, the amount of entanglement generated can be estimated by a
simplified three level model including the state $|h\rangle\equiv
|F',\pm F'\rangle$, for the first/second
ensemble.}\label{Alkalilevels}
\end{center}
\end{figure}
\subsection{Dissipatively driven entanglement between two alkali ensembles}\label{MultilevelCalculation}
In the following, we outline the calculation of the entanglement
which can be produced in the described setting and compute the
time evolution of $\xi_{\text{exp}}(t)$. The master equation
governing the evolution of the atomic system according to the
general model outlined in Sec.~\ref{MultilevelDynamics}, as well
as details of the calculation summed up below, can be found in
App.~\ref{ThreeLevelCalculation}.\\
\\Atomic populations
$N_{\uparrow}\!=\!\sum_{i=1}^{N}|\!\!\uparrow\rangle_{i}\langle
\uparrow\!\!|$,
$N_{\downarrow}\!=\!\sum_{i=1}^{N}|\!\!\downarrow\rangle_{i}\langle
\downarrow \!\!|$ and $N_{h}=\sum_{i=1}^{N}|h\rangle_{i}\langle h
|$ can be calculated using the rate equations
\begin{eqnarray}
&&d_{t}\left(%
\begin{array}{c}
  N_{\uparrow}(t) \\
  N_{\downarrow}(t) \\
  N_{h}(t) \\
\end{array}%
\right)=M\left(%
\begin{array}{c}
  N_{\uparrow}(t) \\
  N_{\downarrow}(t) \\
  N_{h}(t)\\
\end{array}%
\right),\label{Rate equations}
\end{eqnarray}
where
\begin{eqnarray*}
M=\left(%
\begin{array}{ccc}
  -\left(\Gamma_{\uparrow\!\downarrow}+\Gamma_{\uparrow\!h}\right) & \Gamma_{\downarrow\!\uparrow} & \Gamma_{h\!\uparrow} \\
  \Gamma_{\uparrow\!\downarrow} & -\left(\Gamma_{\downarrow\!\uparrow}+\Gamma_{\downarrow\!h}\right) & \Gamma_{h\!\downarrow} \\
   \Gamma_{\uparrow\!h} & \Gamma_{\downarrow\!h} & -2\left(\Gamma_{h\!\uparrow+\Gamma_{h\!\downarrow}}\right) \\
\end{array}%
\right).
\end{eqnarray*}
\begin{figure}[pbt]
\begin{center}
\includegraphics[width=7cm]{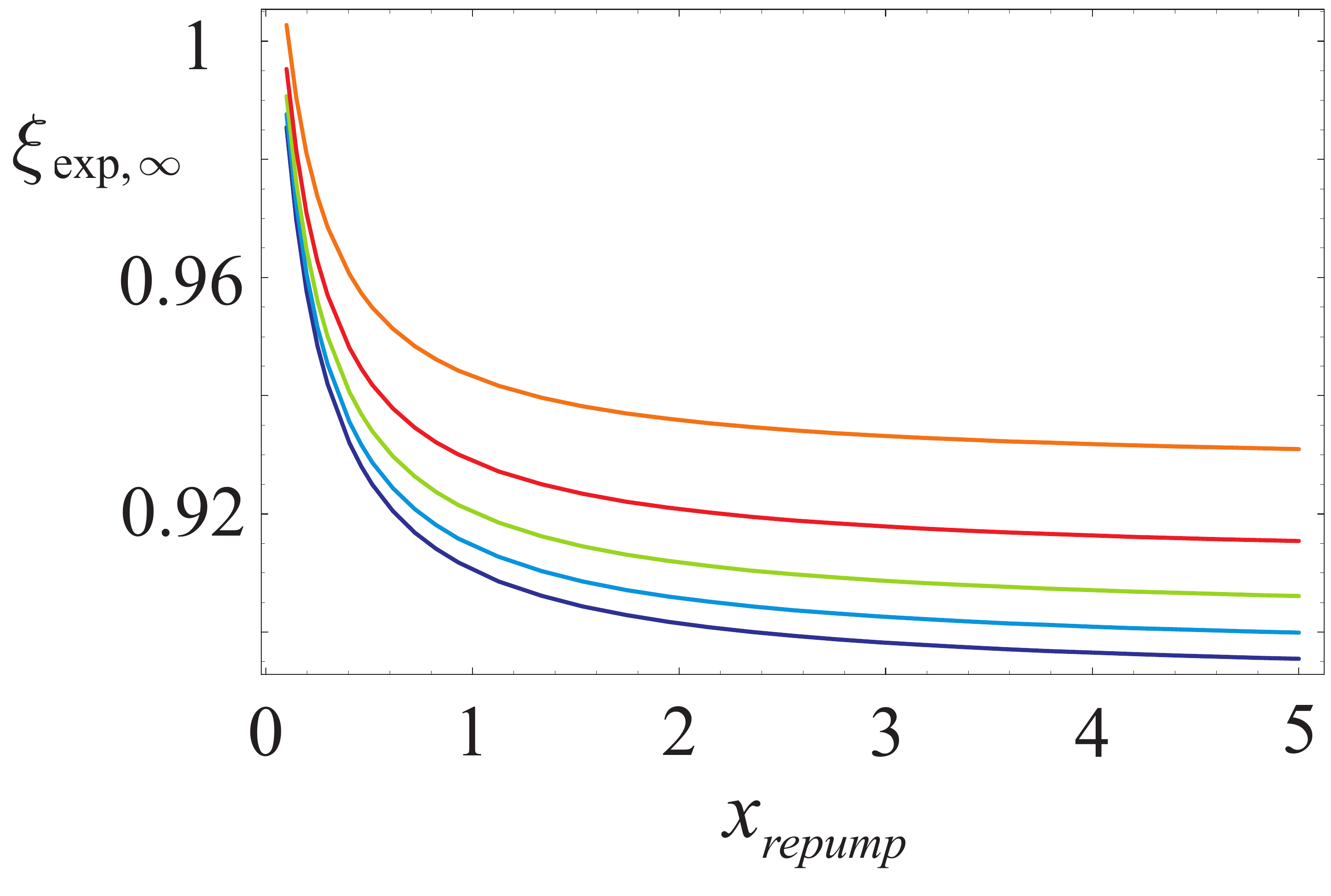}
\caption{(Color online) Steady state entanglement
$\xi_{\text{exp},\infty}$ between two $^{133}\text{Cs}$ ensembles
versus strength of repump fields $x_{\text{repump}}$ (see main
text, Sec.~\ref{MultilevelCalculation} and App.~\ref{Appendix2}).
The plots depict data for ensembles with optical depth $d=30$,
blue detuned probe light with $\Delta=700$MHz and different values
for the non-radiative dephasing rate $\Gamma_{d}^{\text{add}}$.
The curves correspond in ascending order to
$\Gamma_{d}^{\text{add}}=0$ (violet), $\Gamma_{d}^{\text{add}}=2$
(blue), $\Gamma_{d}^{\text{add}}=5$ (green),
$\Gamma_{d}^{\text{add}}=10$ (red) and
$\Gamma_{d}^{\text{add}}=20$ (orange). }\label{CesiumPlot1}
\end{center}
\end{figure}
$\Gamma_{ab}$ is the single-particle rate for the transition
$|a\rangle\rightarrow|b\rangle$. If the transition rates are
known, the number of atoms in the relevant two-level subsystem
$N_{2}(t)=N_{\uparrow}+N_{\downarrow}$ and the polarization
$P_{2}(t)=\left(N_{\uparrow}-N_{\downarrow}\right)/N_{2}(t)$ can
be directly computed.
Based on this result, the time evolution of
$\Sigma_{J,2}=\text{var}\left(J_{y,I}+J_{y,II}\right)_2+\text{var}\left(J_{z,I}-J_{z,II}\right)_2$
can be calculated. Again, the situation exhibits two different
time scales. The decay collective of $\Sigma_{J,2}$ is fast
compared to the evolution of $N_2(t)$ and $P_2(t)$. A calculation
analogous to the one described in App.~\ref{Appendix1} leads to
\begin{eqnarray}\label{3Ent}
\Sigma_{J,2} &=&N_2(0)e^{-\left(\bar{\Gamma}+d(t)\Gamma
P_2(t)\right)t}\\
&+&N_2(t)\frac{\bar{\Gamma}+d(t)\Gamma
P_2(t)^2\left(\mu-\nu\right)^2}{\bar{\Gamma}+d(t)\Gamma P_2(t)}\nonumber\\
&&\left(1-e^{-\left(\bar{\Gamma}+d(t)\Gamma
P_2(t)\right)t}\right),\nonumber
\end{eqnarray}
with $d(t)=d N_2(t)/N$ and
$\bar{\Gamma}=\Gamma_{\uparrow\!\downarrow}+\Gamma_{\downarrow\!\uparrow}+\Gamma_{\uparrow\!h}+\Gamma_{\downarrow\!h}+\Gamma_{\uparrow\!\uparrow}+\Gamma_{\downarrow\!\downarrow}+\Gamma_{\text{d}}^{\text{add}}$,
where $\Gamma_{\text{d}}^{\text{add}}$ accounts for non-radiative
dephasing. On time scales which are long compared to the fast
desired dynamics (but short enough to avoid profuse depletion of
the relevant two-level subsystem, such that $N_2(t)\gg1$ is
guaranteed) $\Sigma_{J,2}$ is given by the long-time (lt) solution
\begin{eqnarray}
\Sigma_{J,2}^{\text{lt}}=N_2(t) \ \!\frac{\bar{\Gamma}+d(t)\Gamma
P_2(t)^2\left(\mu-\nu\right)^2\!}{\bar{\Gamma}+d(t)\Gamma
P_2(t)}.\ \label{var}
\end{eqnarray}
Now, this result is related to experimentally measurable
quantities. Inserting Eqs.~(\ref{Jx}), (\ref{Jy}),
(\ref{TransVar}) and (\ref{var}) into the definition
$\xi_{\text{exp}}=\Sigma_{J,\text{exp}}/ \left(2\mid\!\langle
J_x\rangle_{\text{exp}}\!\mid\right)$ yields
\begin{eqnarray}\label{ExpEnt}
\xi_{\text{exp}}^{\text{lt}}&=&\frac{\bar{\Gamma}+d(t)\Gamma
P_2(t)^2\left(\mu-\nu\right)^2}{\bar{\Gamma}+\!d(t)\Gamma
P_2(t)}\ \!\frac{2F}{P_2(t)+2F-1}\nonumber\\
&+&\frac{N_{\downarrow}(t)}{N_2(t)}\
\!\frac{2(2F-1)}{P_2(t)+2F-1}.\
\end{eqnarray}
This long-time solution is a generalization of
Eq.~(\ref{MainResult}) which takes multi-level dynamics into
account. Particle losses result in a decrease in $d(t)$ and
$P_2(t)$.
If transitions out of the two-level subsystem can be counteracted
efficiently by pump- and repump-fields, a true entangled steady
state can be reached. Else, a quasi steady state is produced.
These two cases are illustrated in Figs. \ref{CesiumPlot1} and
\ref{CesiumPlot2} respectively using the concrete example of
$^{133}\text{Cs}$ ensembles (F=4) at room temperature.
Fig.~\ref{CesiumPlot1} shows the amount of steady state
entanglement generated in case of sufficient repump power. More
specifically, the depicted curves represent solutions for
different values of $\Gamma_{\text{d}}^{\text{add}}$ versus the
repump parameter $x_{\text{repump}}$, starting from
$x_{\text{repump}}=0.01$. The repump parameter $x_{\text{repump}}$
quantifies the strength of the applied repump fields and is given
by the ratio
$x_{\text{repump}}=\Omega^2_{\text{repump}}/\Omega^2_{\text{pump,opt}}$,
where $\Omega^2_{\text{pump,opt}}$ is the optimal Rabi frequency
that can be chosen for the pump field within the validity of the
model considered here. ($\Omega^2_{\text{pump,opt}}$ is the
minimal Rabi frequency of the pump field leading to
 $N_{\uparrow,\infty}/N_{2,\infty}\geq 0.95$ in the steady state.) Details
of the calculations leading to the plots can be found in
App.~\ref{Appendix2}.
Fig.~\ref{CesiumPlot2} illustrates the case, where particle losses
dominate the evolution of $\xi_{\text{exp}}(t)$. The curves in
this figure show the amount of entanglement generated in the
absence of repump fields as a function of time in ms. For short
times the time evolution is governed by the desired dynamics
within the relevant two-level subsystem and reaches quickly an
entangled steady state. For longer times this stable state with
respect to the entangling dynamics is superposed by the slow
additional evolution imposed by the multi-level structure of the
ground state. The fast desired dynamics entangles the collective
spins of both ensembles, while particle losses cause a slow but
continuing shortening of the spins. In this sense,
Fig.~\ref{CesiumPlot2} shows the creation of a quasi steady state.
For $t>0.05/\Gamma$, the time evolution is given by
Eq.~(\ref{ExpEnt}), that is, the time evolution is solely
determined by particle loss-related dynamics.\\
\\In this section, we put emphasis on the general,
implementation-independent limitations of the proposed scheme
imposed by radiative transitions. These undesired processes are
characteristic for a given level scheme and intimately linked to
the tradeoff between enhanced entangling dynamics due to increased
probe,-or pump power and added noise. Depending on the concrete
experimental realization, other undesired processes impairing the
performance of the proposed scheme can occur like for example spin
flips due to collisions. Atomic transitions and additional
dephasing due to collisions with the walls can be taken into
account by including terms of the type
\begin{eqnarray*}
d_t\rho(t)&=&\frac{\Gamma_{\text{col}}}{2}
\sum_{i=1}^{N}\!\!\left(
\sigma_{I,i}\rho(t)\sigma_{I,i}^{\dag}\!+\!
\sigma_{II,i}\rho(t)\sigma_{II,i}^{\dag}\right)\nonumber\\
\!\!&+&\!\!\!\frac{\Gamma_{\rm{col}}}{2}\sum_{i=1}^{N}\!\!\left(
\sigma_{I,i}^{\dag}\rho(t)\sigma_{I,i}\!+\!
\sigma_{II,i}^{\dag}\rho(t)\sigma_{II,i}\right)\nonumber\\
\!\!&+&\!\!\!\frac{\Gamma_{\!\!\rm{col}}}{2}\sum_{i=1}^{N}\left(\sigma_{\downarrow\downarrow,I,i}\rho(t)\sigma_{\downarrow\downarrow,I,i}
+\sigma_{\downarrow\downarrow,II,i}\rho(t)\sigma_{\downarrow\downarrow,II,i}\right)\nonumber\\
&+&...
\end{eqnarray*}
to the master equation. Since the thermal energy of atoms is
typically much larger than the atomic level splittings, one can
assume the same collisional rate $\Gamma_{\text{col}}$ for all
atomic transitions. The value of  $\Gamma_{\text{col}}$ has to be
determined phenomenologically for the specific experimental setup
under consideration (compare \cite{ShortPaper}).
\begin{figure}[pbt]
\begin{center}
\includegraphics[width=7.6cm]{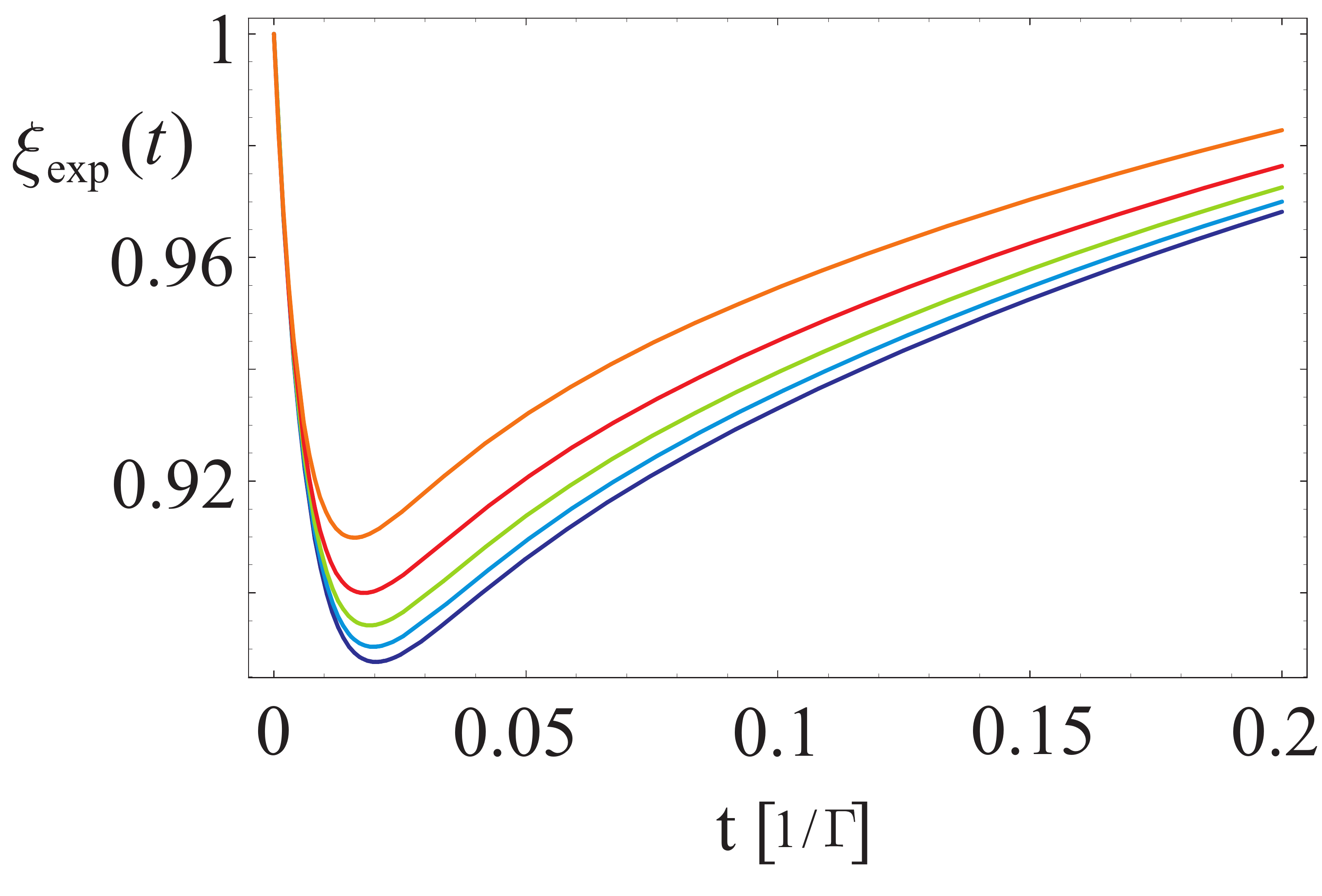}
\caption{(Color online) Quasi steady state entanglement of
$^{133}\text{Cs}$ ensembles versus time in units $1/\Gamma$ in the
absence of repump fields ($x_{\text{repump}}=0$; all other
parameters take values as in Fig.~\ref{CesiumPlot1}). The fast
entanging dynamics results in a drop in $\xi_{\text{exp}}(t)$
(indicating the creation of an inseparable state), but since
particle losses are not counteracted by repump-fields, this
additional slow dynamics eventually causes $\xi_{\text{exp}}(t)$
to rise. Hence, a quasi steady state is produced. The behavior on
long time scales is determined by the stationary state given by
Eq.~(\ref{Main result}) superposed by slow multi-level
dynamics.}\label{CesiumPlot2}
\end{center}
\end{figure}
\section{Conclusions}\label{Conclusions}
In conclusion, we propose a technique for entangling two mesocopic
atomic ensembles at room temperature which are separated by a
macroscopic distance. The core idea is to engineer the coupling of
the atomic system to its environment in such a way that the steady
state of the dissipative time evolution is the desired inseparable
state. As entanglement is produced in the steady state of the
system, it is long-lived and immune to noise.\\
The reservoir consists of the common modes of the electromagnetic
field and the coupling of the bath to the system can be controlled
by means of laser- and magnetic fields. We provide a detailed
theoretical description including dipole-dipole interactions for
two-level systems and find that the imaginary parts of the master
equation (collective Lamb shifts) are negligible. Hence,
light-induced collisions do not play an important role in the
setup considered here. The proposed scheme for generation of
entanglement by dissipation is analyzed for two level systems and
complemented by considering the implementation in multi-level
ground states.\\
\\Future directions include the transfer of the ideas presented
here to other systems, where a quadratic interaction involving two
sideband modes can be realized. The system may either be described
by bosonic modes (compare Sec.~\ref{BasicIdeaCentralResults} and
App.~\ref{Uniqueness}) or spin degrees of freedom. If the
Hamiltonian corresponding to the interaction between this system
and light can be decomposed into an active and a passive part such
that each part involves one sideband, as in
Hamiltonian~(\ref{Hamiltonian}), dissipatively driven entanglement
can be generated using the method described in this paper.\\
\\Optomechanical resonators \cite{Review1,Review2,AspelmeyerReview} interacting with light are promising
candidates. If an optomechanical system is driven by a strong pump
laser, the linearized radiation-pressure Hamiltonian gives rise to
a passive (beam-splitter) interaction for positive detuning
between cavity and pump-frequency  \cite{Linearization1,
Linearization2}. The resulting effective interaction is active
(squeezing) for negative detuning. The effective optomechanical
coupling rates can be adjusted by tuning the intensities of the
driving fields \cite{CouplingRate}. Hence, the quadratic
interaction in this system provides naturally the basic
prerequisites for implementation of the proposed scheme.
One can for example envision a setup, where light interacts
subsequently with two movable mirrors which each constitute one
end of an optical cavity such that the nanomechanical oscillators
are driven into an entangled state. Equivalently, membranes
coupled to Fabri-Perot cavity modes could be used.
\subsection*{Acknowledgements}
We thank Geza Giedke, Karl Gerd Vollbrecht, Klemens Hammerer,
Markus Aspelmeyer and Diego Porras for very valuable discussions
and express our gratitude to Hanna Krauter, Kasper Jensen and
Wojciech Wasilewski for their input concerning the implementation
of the proposed scheme in Cesium vapor cells. We acknowledge
support from the Elite Network of Bavaria (ENB) project QCCC, the
DFG-Forschungsgruppe 635 and the EU projects COMPAS, Q-ESSENCE and
QUEVADIS.
\appendix
\section{Steady state entanglement for bosonic modes}\label{Uniqueness}
In the following, it is shown that the two mode squeezed state
$\rho_{\text{TMS}}=|\Psi_{\text{TMS}}\rangle\langle
\Psi_{\text{TMS}}|$, with
$\tilde{A}|\Psi_{\text{TMS}}\rangle=\tilde{B}|\Psi_{\text{TMS}}\rangle=0$,
is the unique steady state of the time evolution described by the
Master equation
\begin{eqnarray*}
d_{t}\rho(t)&=&\kappa_{A} \left(\tilde{A} \rho(t)
\tilde{A}^{\dag}-\tilde{A}^{\dag} \tilde{A}
\rho(t)/2-\rho(t)\tilde{A}^{\dag}\tilde{A}/2\right)\nonumber\\
&+& \kappa_{B}\left( \tilde{B} \rho(t)
\tilde{B}^{\dag}-\tilde{B}^{\dag} \tilde{B}
\rho(t)/2-\rho(t)\tilde{B}^{\dag}\tilde{B}/2\right).
\end{eqnarray*}
As stated in Sec.~\ref{BasicIdeaCentralResults}, the bosonic mode
operators $a$ and $b$ ($[a,a^{\dag}]=[b,b^{\dag}]=1$) can be
transformed into the non-local operators $\tilde{A}$ and
$\tilde{B}$ by the unitary operation
\begin{eqnarray*}
    \tilde{A}&=&UaU^{\dag}=\mu\  a+\nu \ b^{\dag},\\
    \tilde{B}&=&UbU^{\dag}=\mu\  b+\nu \ a^{\dag}.
\end{eqnarray*}
Since unitary transformations preserve commutation relations,
$[\tilde{A},\tilde{A}^{\dag}]=[\tilde{B},\tilde{B}^{\dag}]=1$ with
$\mu^2-\nu^2=1$.
By inserting these expressions in the equation above and defining
$\rho_{U}=U^{\dag}\rho U$ we find
\begin{eqnarray}
d_{t}\rho_{U}(t)\!\!\!&=&\!\!\!\kappa_{A}\! \left(a \rho_{U}(t)
a^{\dag}\!-\!a^{\dag} a
\rho_U(t)/2\!-\!\rho_U(t)a^{\dag}a/2\right)\nonumber\\
\!\!\!&+&\!\!\! \kappa_{B}\!\left(b \rho_U(t)
b^{\dag}\!-\!b^{\dag}
b\rho_U(t)/2\!-\!\rho_U(t)b^{\dag}b/2\right)\!\!.\label{ZeroTemperatureBath}
\end{eqnarray}
This is a master equation for two modes coupled to a bath with
temperature $T=0$. The steady state is the vacuum
$|0,0\rangle\langle 0,0|$, with $a|0,0\rangle=b|0,0\rangle=0$
\cite{Vacuum}.
Hence, inverting the unitary transformation yields the unique
steady state $U|0,0\rangle=|\Psi_{\text{TMS}}\rangle$.\\
\\For bosonic modes, the amount of entanglement can be quantified
by means of the violation of a local uncertainty relation in terms
of quadratures \cite{EPR1, EPR2}. For entangled states
\begin{eqnarray*}
\text{var}(x_+)+\text{var}(p_-)<1,
\end{eqnarray*}
where $x_{+}=(x_{\text{a}}+x_{\text{b}})/\sqrt{2}$,
$p_{-}=(p_{\text{a}}-p_{\text{b}})/\sqrt{2}$ and
$x_{\text{a}}=\left(a+a^\dag\right)/ \sqrt{2}$ and
$p_{\text{a}}=-i\left(a-a^\dag\right)/ \sqrt{2}$ (analogous
expressions hold for $x_{\text{b}}$ and $p_{\text{b}}$). In
particular,
$\text{var}(x_+)+\text{var}(p_-)=\left(\mu-\nu\right)^{-2}=e^{-2r}$,
for two mode squeezed states with squeezing parameter $r$.\\
\\For large, strongly polarized atomic ensembles, collective spins
can be described by bosonic modes
$\frac{1}{\sqrt{N_I}}\sum_{i=1}^{N_{I}}\sigma_{I,i}\approx a$,
$\frac{1}{\sqrt{N_{II}}}\sum_{i=1}^{N_{II}}\sigma_{II,i}\approx b$
using the Holstein-Primakoff-approximation
\cite{Holstein-Primakoff}. In this case, $\xi<1$ (see
Sec.~\ref{BasicIdeaCentralResults}) is equivalent to the criterion
$\text{var}(x_+)+\text{var}(p_-)<1$.
\section{Derivation of the master equation}\label{DerivationME}
%
%
%
In this appendix, the master equation for creating entanglement
between two atomic ensembles at room temperature discussed in
Sec.~\ref{TwoLevelModel}, is derived in detail.
App.~\ref{DerivationME-I} and App.~\ref{DerivationME-II}
complement  Sec.~\ref{MasterEquation2} and
Sec.~\ref{MasterEquation3} respectively. In the former, we derive
the master equation for atomic ground states,
Eq.~(\ref{MEafterAdiabaticElimination}). In the latter, we include
thermal motion of atoms and calculate the effective decay rates.
We show that in the setup under consideration, the resulting
master equation can be assumed to be real, since imaginary parts
play only a minor role.
\subsection{Master equation for atomic ground state levels $|\!\!\uparrow\rangle$ and $|\!\!\downarrow\rangle$}\label{DerivationME-I}
%
%
%
Light and matter are assumed to interact as described in
Sec.~\ref{MasterEquation1}. We consider the full Hamiltonian
including undesired transitions \cite{UndesiredTransitions} and
without applying the rotating wave approximation for quantum
fields. It is given by
$H=H_{\text{L}}+H_{\text{A}}+H_{\text{int}}$, where
$H_{\text{L}}=\int dk\
\left(\omega_k-\omega_{L}\right)a_k^{\dag}a_k$ is the Hamiltonian
of the free light field and $H_{\text{A}}$ accounts for atomic
energies in the rotating frame.
$H_{\text{A}}=H_{\text{A,I}}+H_{\text{A,II}}$ with
$H_{\text{A,I}}=\sum_{i}\left(\Delta_{\uparrow,I}|e_{\uparrow}\rangle_{I,i}\langle
e_{\uparrow}|+\Delta_{\downarrow,I}|e_{\downarrow}\rangle_{I,i}\langle
e_{\downarrow}|\right)$ and
$H_{\text{A,II}}=\sum_{i}\left(\Delta_{\uparrow,II}|e_{\uparrow}\rangle_{II,i}\langle
e_{\uparrow}|+\Delta_{\downarrow,II}|e_{\downarrow}\rangle_{II,i}\langle
e_{\downarrow}|\right)$. Here, we introduced the detunings
$\Delta_{\uparrow,I/II}$ and $\Delta_{\downarrow,I/II}$ which
correspond to diagonal transitions
$|\!\!\downarrow\rangle\rightarrow|e_{\uparrow}\rangle$ and
$|\!\!\uparrow\rangle\rightarrow|e_{\downarrow}\rangle$
respectively in the first/second ensemble. In the setup
illustrated in Fig.~\ref{Figure2},
$\Delta_{\uparrow,I}=-\Delta_{\uparrow,II}=\Delta-\Omega$ and
$\Delta_{\downarrow,I}=-\Delta_{\downarrow,II}=\Delta+\Omega$. The
interaction Hamiltonian
$H_{\text{int}}=H_{\text{cl}}+H_{\text{qu}}$ consists of a
classical part $H_{\text{cl}}$, which accounts for transitions
induced by the driving field and a quantum part $H_{\text{qu}}$,
which involves quantized field operators. The Hamiltonian
describing the interaction of light with the first atomic ensemble
is governed by the Hamiltonian
\begin{eqnarray}\label{FullHamiltonian}
H_{\text{int,I}}\!\!&=&\!\!H_{\text{cl,I}}+H_{\text{qu,I}},\sigma_{e_{\uparrow}\uparrow,I,i}\\
H_{\text{cl,I}}\!\!&=&\!\!\Omega_{\text{probe}}\sum_{i=1}^{N}e^{i\mathbf{k}_L \mathbf{r}_i}\left(|e_{\uparrow}\rangle_{\!I,i}\langle \downarrow\!\!|+|e_{\downarrow}\rangle_{\!I,i}\langle \uparrow\!\!|\right)+H.C.,\nonumber\\
H_{\text{qu,I}}\!\!&=&\!\!\!\sum_{i=1}^N\sum_{\mathbf{k}}\!\sum_{\hat{\lambda}_{\mathbf{k}}=1}^{2}\!
g_\mathbf{k} e^{i
\mathbf{k}\mathbf{r}_i}a_{\mathbf{k}}(\!|e_{\downarrow}\rangle_{\!I,i}\langle
\downarrow\!\!|e^{i\Omega t}\!+\!|e_{\uparrow}\rangle_{\!I,i}\langle \uparrow\!\!|e^{-i\Omega t}\nonumber\\
\!\!&+&\!\!\!|\!\!\downarrow\rangle_{\!I,i}\langle
e_{\downarrow}|e^{-i\Omega t}\!e^{-2i\omega_L
t}\!+\!|\!\!\uparrow\rangle_{\!I,i}\langle e_{\uparrow}|e^{i\Omega
t}\!e^{-2i\omega_L t})\!+\!H.C.,\nonumber\\
&+&\!\!\!\sum_{i=1}^N\sum_{\mathbf{k}}\!
\sum_{\hat{\lambda}_{\mathbf{k}}=1}^{2}\! \check{g}_\mathbf{k}e^{i
\mathbf{k}\mathbf{r}_i}a_\mathbf{k}(|e_{\uparrow}\rangle_{\!I,i}\langle
\downarrow\!\!|\!+|e_{\downarrow}\rangle_{\!I,i}\langle \uparrow\!\!|\nonumber\\
\!\!&+&\!\!\!|\!\!\downarrow\rangle_{\!I,i}\langle
e_{\uparrow}|e^{-2i\omega_L
t}\!+\!|\!\!\uparrow\rangle_{\!I,i}\langle
e_{\downarrow}|e^{-2i\omega_L t})\!+\!H.C.,\nonumber
\end{eqnarray}
where $\hat{\lambda}_{\mathbf{k}}$ specifies the two orthogonal
polarizations of the light mode with wave vector $\mathbf{k}$.
$g_{\mathbf{k}}=\hat{\mathbf{e}}_{\mathbf{k}}
\cdot\mathbf{p}\sqrt{\frac{\omega_{k}}{2\epsilon_{0} V}}$ is the
coupling strength of desired transitions involving the quantum
field, while $\check{g}_{\mathbf{k}}$ is the coupling strength
corresponding to undesired transitions.
$\hat{\mathbf{e}}_{\mathbf{k}}$ is the unit polarization vector,
$V$ is the quantization volume of the electromagnetic field,
$\epsilon_0$ the vacuum permittivity and $\mathbf{p}$ is the
transition matrix element of transitions
$|e_{\uparrow}\rangle\rightarrow |\downarrow\rangle$ and
$|e_{\downarrow}\rangle\rightarrow |\uparrow\rangle$.
$H_{\text{int}}=H_{\text{int,I}}+H_{\text{int,II}}$.
$H_{\text{int,II}}$ is given by an expression analogous to
Eq.~(\ref{FullHamiltonian}), where subscripts are changed
accordingly ($I \rightarrow II$).\\
%
%
%
\\Based on this Hamiltonian, we derive a master equation for the
reduced atomic density matrix $\rho(t)$ using the approximation of
independent rates of variation and assuming Born-Markov dynamics
as explained in Sec.~\ref{MasterEquation2}
\begin{eqnarray*}
d_t\rho(t)=-i[H_{cl},\rho(t)]+{\cal L}_{\text{qu}}\rho(t),
\end{eqnarray*}
where $\cal{L}$ is a  Lindblad operator corresponding to the
quantum part of the interaction $H_{\text{qu}}$. This master
equation consists of three parts
\begin{eqnarray}\label{ThreeParts}
d_t\rho(t)\!\!=\!\!L(\rho(t))_{\text{ens.I}}\!+\!L(\rho(t))_{\text{ens.II}}\!+\!L(\rho(t))_{\text{inter-ens.}}\
.
\end{eqnarray}
The first and second term $L(\rho(t))_{\text{ens.I}}$ and
$L(\rho(t))_{\text{ens.II}}$ include only terms referring to the
first and second ensemble respectively, while the third term
$L(\rho(t))_{\text{inter-ens.}}$ summarizes all terms combining
operators acting on both samples. The first term is given by
\begin{eqnarray}\label{L1/2}
L(\rho(t))_{\text{ens.I}}&=&-i[H_{cl,I},\rho(t)]\\
&+&\frac{1}{2}\sum_{i,j}\mathcal{J}^{I,I}_{ij}|\!\!\uparrow\rangle_{\!I,i}\langle
e_{\uparrow}|\rho(t)|e_{\uparrow}\rangle_{\!I,j}\langle
\uparrow \!\!|\nonumber\\
&+&\frac{1}{2}\sum_{i,j}\check{\mathcal{J}}^{I,I}_{ij}|\!\!\downarrow\rangle_{\!I,i}\langle
e_{\uparrow}|\rho(t)|e_{\uparrow}\rangle_{\!I,j}\langle
\downarrow \!\!|\nonumber\\
&+&...\ ,\nonumber
\end{eqnarray}
where we used the approximation $\omega_{L}\gg \Omega$, which is
very well justified for optical frequencies, and neglected fast
oscillating terms which appear in the standard derivation
\cite{CT}, if photonic modes in the upper and lower sideband are
not treated as independent baths \cite{IndependentBaths}.
$\mathcal{J}^{I,I}_{ij}=\mathcal{J}^{I,I}_{ji}$ is a complex decay
rate associated with desired transitions, while
$\check{\mathcal{J}}^{I,I}_{ij}=\check{\mathcal{J}}^{I,I}_{ji}$ is
the rate of undesired transitions. For the simple model discussed
in Sec.\ref{TwoLevelModel},
$\check{\mathcal{J}}^{I,I}_{ij}=2\mathcal{J}^{I,I}_{ij}$.
Imaginary single particle terms
$Im\left(\mathcal{J}^{I,I}_{ii}\right)$ (Lamb shifts) are absorbed
in the detunings as explained in Sec.~\ref{MasterEquation2}.
The second term $L(\rho(t))_{\text{ens.II}}$ is given by an
analogous expression with changed subscripts $I\mapsto II$.\\
\\The last term in Eq.~(\ref{ThreeParts}) can be written as sum
$L(\rho(t))_{\text{inter-ens.}}=L(\rho(t))_{\text{inter-ens.}}^{I,II}+L(\rho(t))_{\text{inter-ens.}}^{II,I}$.
The first part is given by
\begin{eqnarray}\label{L3}
L(\rho(t))_{\text{inter-ens.}}^{I,II}&=&\frac{1}{2}\sum_{i,j}\mathcal{J}^{I,II}_{ij}|\uparrow\rangle_{I,i}\langle
e_{\uparrow}|\rho(t)|e_{\downarrow}\rangle_{II,j}\langle \downarrow\!\!|\nonumber\\
&+&\frac{1}{2}\sum_{i,j}\mathcal{J}^{I,II}_{ij}|\downarrow\rangle_{I,i}\langle
e_{\downarrow}|\rho(t)|e_{\uparrow}\rangle_{II,j}\langle \uparrow \!\!|\nonumber\\
&+&\frac{1}{2}\sum_{i,j}\check{\mathcal{J}}^{I,II}_{ij}|\uparrow\rangle_{I,i}\langle
e_{\downarrow}|\rho(t)|e_{\uparrow}\rangle_{II,j}\langle \downarrow \!\!|\nonumber\\
&+&\frac{1}{2}\sum_{i,j}\check{\mathcal{J}}^{I,II}_{ij}|\downarrow\rangle_{I,i}\langle
e_{\uparrow}|\rho(t)|e_{\downarrow}\rangle_{II,j}\langle \uparrow \!\!|\nonumber\\
&+&...
\end{eqnarray}
and $L(\rho(t))_{\text{inter-ens.}}^{II,I}$ is given by an
analogous expression with changed subscripts ($I\rightarrow II$).
Decay rates appearing in inter-ensemble terms differ from
single-ensemble rates
$\mathcal{J}^{I,I}_{ij}=\mathcal{J}^{II,II}_{ij}$
\begin{eqnarray*}
\mathcal{J}_{ij}^{I,II}=e^{-i\mathbf{k}\mathbf{R}}\mathcal{J}^{I,I}_{ij},\
\mathcal{J}_{ij}^{II,I}=e^{+i\mathbf{k}\mathbf{R}}\mathcal{J}^{I,I}_{ij}
\end{eqnarray*}
where $\mathbf{R}$ is the distance between the two ensembles. In
order to obtain compact expressions, we use the simplified
notation $\mathcal{J}_{ij}$ for single ensemble or inter-ensemble
rates depending to which samples the indices $i$ and $j$ refer. We
use the convention
$\mathbf{r}_{ij}=\mathbf{r}_i-\mathbf{r}_j-\mathbf{R}$ if the atom
with index $j$ is located in the second ensemble. (However, in
App.~\ref{DerivationME-II}, it is shown that the distance between
the ensembles does not play a role in the setting considered
here.) Using this notation, $\mathcal{}\mathcal{J}_{ij}$ is given
by
\begin{eqnarray*}
\mathcal{J}_{ij}\!\!&=&\!\!\!\!\int\!\!\! d\mathbf{k}
\!\!\sum_{\hat{\lambda}_{\mathbf{k}}=1}^{2}\!\!g\!^{2}(\mathbf{k})e^{i\mathbf{k}(\mathbf{r}_i-\mathbf{r}_j)}\!\!\!\int_{0}^{\infty}
\!\!\!\!\!d\tau
\!(\!e^{-i(\omega_{k}\!-\omega_{L})\tau}\!\!+\!e^{-i(\omega_{k}\!+\omega_{L})\tau}\!)\!,
\end{eqnarray*}
where the sum over light modes was changed into an integral
$\sum_{\mathbf{k}}\rightarrow\frac{V}{\left(2\pi\right)^3}\int
{d_{\mathbf{k}}}$. The prefactor is absorbed in the coupling
constant $g(\mathbf{k})=\sqrt{V/(2\pi)^3}\ g_{\mathbf{k}}$
whenever an integral over light modes is used. The second term in
the expression in brackets ($e^{-i(\omega_{k}+\omega_{L})\tau}$)
stems from counter-rotating terms in the Hamiltonian, and would
not appear if the rotating wave approximation had been applied.
Using the identity $\int_0^{\infty}e^{i\omega
\tau}=\pi\delta(\omega)+i\mathcal{P}(1/\omega)$, where $\cal{P}$
is the principal value, we obtain
\begin{eqnarray*}
Re\!\left(\mathcal{J}_{ij}\right)\!\!\!&=&\!\!\!\pi\!\!\int\!\!\!
d\mathbf{k}\!\!\sum_{\hat{\lambda}_{\mathbf{k}}=1}^{2}\!g^{2}(\mathbf{k})e^{i\mathbf{k}(\mathbf{r}_i-\mathbf{r}_j)}\delta(\omega_k-\omega_L),\\
Im\!\left(\mathcal{J}_{ij}\right)\!\!\!&=&\!\!\!i\mathcal{P}\!\!\left(\!\int\!\!\!
d\mathbf{k}\!\!\sum_{\hat{\lambda}_{\mathbf{k}}=1}^{2}\!\!g^{2}\!(\mathbf{k})e^{i\mathbf{k}(\mathbf{r}_i-\mathbf{r}_j)}\!\!\left(\!
\frac{1}{\omega_L-\omega_k}\!+
\!\frac{1}{\omega_L+\omega_k}\!\right)\!\!\right)\!\!,
\end{eqnarray*}
These rates can be calculated as shown in \cite{Lehmberg} (compare
Eqs.~(\ref{RealPart}) and (\ref{ImaginaryPart})).\\
\\Now, a master equation for the
reduced density matrix of atomic ground states
$\mathbb{P}_g\rho(t)\mathbb{P}_g$ is derived by applying the
projector
$\mathbb{P}_g=\bigotimes_{i,j=1}^{N}\left(|\!\!\uparrow\rangle_{I,i}\langle
\uparrow\!\!|+|\!\!\downarrow\rangle_{I,i}\langle
\downarrow\!\!|\right)\otimes
\left(|\!\!\uparrow\rangle_{II,j}\langle
\uparrow\!\!|+|\!\!\downarrow\rangle_{II,j}\langle
\downarrow\!\!|\right)$ to the differential equation
$d_t\rho(t)=L(\rho(t))_{\text{ens.I}}+L(\rho(t))_{\text{ens.II}}+L(\rho(t))_{\text{inter-ens.}}$
using Eqs. (\ref{L1/2}) and (\ref{L3}). Excited states are
eliminated under the condition
$\Delta_{\uparrow,I/II},\Delta_{\downarrow,I/II}>>\Gamma_{\text{atomic}}$,
using
$d_t\mathbb{P}_e\rho(t)\mathbb{P}_g=d_t\mathbb{P}_g\rho(t)\mathbb{P}_e=d_t\mathbb{P}_e\rho(t)\mathbb{P}_e=0$,
where $\mathbb{P}_e=1\!\text{I}-\mathbb{P}_g$.
Moreover, we assume that terms corresponding to states with two or
more excitations, for example terms of the type
$\mathbb{P}_e^{(2)}\rho(t)\mathbb{P}_g$, are negligible compared
to terms corresponding to states where at most one atom is in an
excited state like $\mathbb{P}_e^{(1)}\rho(t)\mathbb{P}_g$.
$\mathbb{P}^{(1)}_{e}=\sum_{i=1}^{N}\mathbb{P}_{e,I,i}\otimes\mathbb{P}_{g,II}+\sum_{j=1}^{N}\mathbb{P}_{g,I}\otimes\mathbb{P}_{e,II,j}$
with $\mathbb{P}_{e,i,I/II}= \bigotimes_{i=1}^{N}
\left(|e_{\uparrow}\rangle_{I/II,i}\langle
e_{\uparrow}|+|e_{\downarrow}\rangle_{I/II,i}\langle
e_{\downarrow}|\right)$. $\mathbb{P}_e^{(2)}$ is defined
analogously.
In the following we denote the resulting reduced density matrix of
atomic ground states $\mathbb{P}_g\rho(t)\mathbb{P}_g$ simply by
$\rho(t)$. We obtain
\begin{eqnarray*}
d_t\rho(t)&=&\frac{\Omega_{\text{probe}}}{2}\sum_{i,j}e^{-i\mathbf{k}_L\left(r_j-r_i\right)}(\\
\!\!\!&&\!\!\!\mathcal{J}_{ij}\!\!\left(\!\frac{\sigma_{I,i}}{\Delta_{\uparrow,I}}\!+\!\frac{\sigma_{II,j}^{\dag}}{\Delta_{\downarrow,II}}\!\right)\!\!
\rho(t)\!\!\left(\!\frac{\sigma_{I,i}^{\dag}}{\Delta_{\uparrow,I}}\!+\!\frac{\sigma_{II,j}}{\Delta_{\downarrow,II}}\!\right)\\
\!\!\!&+&\!\!\!\mathcal{J}_{ij}\!\!\left(\!\frac{\sigma_{II,i}}{\Delta_{\uparrow,II}}\!+\!\frac{\sigma_{I,j}^{\dag}}{\Delta_{\downarrow,I}}\right)\!\!
\rho(t)\!\!\left(\!\frac{\sigma_{II,i}^{\dag}}{\Delta_{\uparrow,II}}\!+\!\frac{\sigma_{I,j}}{\Delta_{\downarrow,I}}\!\right)\\
\!\!\!&+&\!\!\!\check{\mathcal{J}}_{ij}\!\!\left(\!\frac{\sigma_{\downarrow\downarrow,I,i}}{\Delta_{\uparrow,I}}\!+\!\frac{\sigma_{\uparrow\uparrow,II,j}^{\dag}}{\Delta_{\downarrow,II}}\!\right)\!\!
\rho(t)\!\!\left(\frac{\sigma_{\downarrow\downarrow,I,i}}{\Delta_{\uparrow,I}}\!+\!\frac{\sigma_{\uparrow\uparrow,II,j}^{\dag}}{\Delta_{\downarrow,II}}\right)\\
\!\!\!&+&\!\!\!\check{\mathcal{J}}_{ij}\!\!\left(\!\frac{\sigma_{\downarrow\downarrow,II,i}}{\Delta_{\uparrow,II}}\!+\!\frac{\sigma_{\uparrow\uparrow,I,j}^{\dag}}{\Delta_{\downarrow,I}}\!\right)\!\!
\rho(t)\!\!\left(\frac{\sigma_{\downarrow\downarrow,II,i}}{\Delta_{\uparrow,II}}\!+\!\frac{\sigma_{\uparrow\uparrow,I,j}^{\dag}}{\Delta_{\downarrow,I}}\right))\\
&+&...\ ,
\end{eqnarray*}
where the abbreviations
$\sigma_{\uparrow\uparrow,I/II,i}=|\uparrow\rangle_{I/II,i}\langle
\uparrow|$ and
$\sigma_{\downarrow\downarrow,I/II,i}=|\downarrow\rangle_{I/II,i}\langle
\downarrow|$ were used. Terms with prefactors $1/\Delta^3$ have
been neglected since we assume that detunings are large. AC-Stark
shifts
\begin{eqnarray*}
d_t\rho(t)|_{\text{AC Stark}}&=&-i\Omega_{\text{probe}}\sum_{i=1}^N\left[\frac{\sigma_{\uparrow\uparrow,I,i}}{\Delta_{\downarrow,I}}+\frac{\sigma_{\downarrow\downarrow,I,i}}{\Delta_{\uparrow,I}},\ \rho(t)\right]\\
&+&i\Omega_{\text{probe}}\sum_{i=1}^N\left[\frac{\sigma_{\uparrow\uparrow,II,i}}{\Delta_{\downarrow,I}}+\frac{\sigma_{\downarrow\downarrow,II,i}}{\Delta_{\uparrow,I}},\
\rho(t)\right]
\end{eqnarray*}
are absorbed in the detunings. Using the definitions
$\mu_{I/II}=\pm\frac{\Delta +\Omega}{2\sqrt{\Delta \Omega}}$,
$\nu_{I/II}=\pm\frac{\Delta -\Omega}{2\sqrt{\Delta \Omega}}$
\cite{Normalization} and $J_{ij}=\mathcal{J}_{ij}\
2\Omega_{\text{probe}}\sqrt{\Delta
\Omega}/\left(\Delta^2-\Omega^2\right)$, one obtains
\begin{eqnarray}
d_{t}\rho(t)\!\!\!&=&\!\!\!\frac{1}{2}\sum_{i,j=1}^N e^{-i
\mathbf{k}_{L}\left(\mathbf{r}_{j}-\mathbf{r}_i\right)}J_{ij}\left(
A_i\rho(t) A_{j}^{\dag} + B_i\rho(t)
B_{j}^{\dag}\right)\nonumber\\
\!\!\!&+&\!\!\!\frac{1}{2}\sum_{i,j=1}^N e^{-i
\mathbf{k}_{L}\left(\mathbf{r}_{j}-\mathbf{r}_i\right)}\check{J}_{ij}\left(C_i\rho(t)
C_{j}^{\dag}+D_i\rho(t) D_{j}^{\dag}\right)\nonumber\\
\!\!\!&+&\!\!\!... \label{MEafterAdiabaticElimination}
\end{eqnarray}
with
\begin{eqnarray*}
A_i&=&\mu_I \ \sigma_{I,i}+\nu_{II}\ \sigma^{\dag}_{II,i},\\
B_i&=&\mu_{II}\  \sigma_{II,i}+\nu_{I}\ \sigma^{\dag}_{I,i},\nonumber\\
C_i&=&\mu_{I}\  \sigma_{\downarrow\downarrow,I,i}+\nu_{II}\ \sigma_{\uparrow\uparrow,II,i},\nonumber\\
D_i&=&\mu_{II}\ \sigma_{\downarrow\downarrow,II,i}+\nu_{I}\
\sigma_{\uparrow\uparrow,I,i}.\nonumber
\end{eqnarray*}
The expressions in the main text are obtained by introducing a
unified notation (compare Eq.~(\ref{Operators})), which absorbs
the relative sign $ \mu_{I}/\mu_{II}$ in atomic operators
referring to the second ensemble
$\sigma_{II}\rightarrow\text{sgn}(\mu_{I}\mu{II})\sigma_{II}$.
\subsection{Master equation including atomic motion}\label{DerivationME-II}
Below, we include thermal motion of particles, by treating atomic
positions as classical random variables.\\
We start from the master equation for atomic ground states
(\ref{MEafterAdiabaticElimination}). As outlined in
Sec.~\ref{MasterEquation3}, random atomic positions can be taken
into account by introducing averaged coefficients in the master
equation. Averaged rates $\langle J_{ij}\rangle$ are calculated
assuming Gaussian distributions with width $L$ for atomic
positions in the two ensembles. We start by calculating the rate
corresponding to moving particles in a single ensemble. Below it
is shown that for inter-ensemble rates the same result is obtained
for the setup and range of parameters considered here.
\begin{eqnarray}\label{AveragedRate}
\langle J_{ij}\rangle\!\!\!&=&\!\!\!\frac{1}{\pi^3L^6}\!\int
\!\!d\mathbf{r}\!\!\int\!\! d\mathbf{\acute{r}}
e^{i\mathbf{k}_L\left(\mathbf{r}-\mathbf{\acute{r}}\right)-\frac{r^2+\acute{r}^2}{L^2}}
\!\left(\gamma\!\left(\mathbf{r}-\mathbf{\acute{r}}\right)\!+\!ig\!\left(\mathbf{r}-\mathbf{\acute{r}}\!\right)\!\right)\!,\nonumber\\
\!\!\!&=&\!\!\!\frac{1}{\left(2\pi\right)^{3/2}L^3}\int d
\mathbf{r}_-e^{i\mathbf{k}_L\mathbf{r_-}-\frac{\mathbf{r_-}^2}{2L^2}}\!\left(\gamma\!\left(\mathbf{r_-}\right)\!+\!ig\!\left(\mathbf{r_-}\!\right)\!\right)\!,
\end{eqnarray}
where the variable transformation
$\mathbf{r_{+}}=\mathbf{r}+\mathbf{\acute{r}}$,
$\mathbf{r_{-}}=\mathbf{r}-\mathbf{\acute{r}}$ was made in the
second step. The single particle rate is given by  $J_{ii}=\Gamma$
(Lamb shifts are absorbed in the detunings). Now, averaged rates
$\langle J_{ij}\rangle$ with $i \neq j$ need to be determined.
We consider now first the real part and then the imaginary part of
the averaged decay rate $\langle J_{ij}\rangle$.\\
\\The real part $\Gamma_{ij}=Re\left(\langle J_{ij}\rangle\right)$ is calculated by inserting Eq.~(\ref{RealPart})
into Eq.~(\ref{AveragedRate}) and fixing
$\mathbf{\hat{p}}=\mathbf{\hat{x}}$ and
$\mathbf{\hat{k}_L}=\mathbf{\hat{z}}$.
$\Gamma_{ij}=\Gamma_{ij,A}+\Gamma_{ij,B}$ is a sum of two
contributions corresponding to the first and the second line in
Eq.~(\ref{RealPart}). The first term is given by
\begin{eqnarray}\label{GammaA}
\Gamma_{ij,A}&=&\frac{3\Gamma}{\sqrt{2 \pi}L^3}\int_{0}^\infty
dr_- r_-^2 e^{-\frac{r_-^2}{2L^2}}\
\frac{\sin(k_L r_-)}{k_L r_-}\\
&&\left(\frac{\sin(k_L r_-)}{k_L r_-}+\frac{\cos(k_L
r_-)}{\left(k_L r_-\right)^2}-\frac{\sin(k_L r_-)}{\left(k_L
r_-\right)^3}\right).\nonumber
\end{eqnarray}
The integrand of this expression tends to zero in the limit
$r_-\rightarrow 0$. This can for example be seen by expanding the
integrand for small values $r_-\ll 1$, $r_-^2
e^{-\frac{r_-^2}{2L^2}}\ \frac{\sin(k_L r_-)}{k_L r_-}
\left(\frac{\sin(k_L r_-)}{k_L r_-}+\frac{\cos(k_L r_-)}{\left(k_L
r_-\right)^2}-\frac{\sin(k_L r_-)}{\left(k_L
r_-\right)^3}\right)=\frac{2}{3}r_-^2+\mathcal{O}(r_-^4)$. Hence
the dominant contribution in the limit $k_{L}L\gg 1$, stems from
the first term in brackets $\sin(k_L r_-)/(k_L r_-)$. The other
two terms in brackets decay faster in the interatomic distance
$r_-$ and lead only to corrections on the order of $1/(k_L L)^4$.
For $k_L L\gg 1$,
\begin{eqnarray*}
\Gamma_{ij,A}&=&\frac{3\Gamma}{\sqrt{2 \pi}L^3}\int_{0}^\infty
dr_- r_-^2 e^{-\frac{r_-^2}{2L^2}}\ \frac{\sin(k_L
r_-)^2}{\left(k_L r_-\right)^2}\\
&=&\frac{3}{4}\Gamma\frac{1}{\left(k_{L}L\right)^2}\left(1-e^{-2
k_L L}\right),
\end{eqnarray*}
which can be approximated by
$\Gamma_{ij,A}=\frac{3}{4}\Gamma\frac{1}{\left(k_{L}L\right)^2}$.
The second part
\begin{eqnarray*}
\Gamma_{ij,B}\!\!\!&=&\!\!\!\frac{3 \Gamma}{\sqrt{2
\pi}L^3}\!\int_{0}^\infty\!\!\!\!\!dr_-r_-^2
e^{-\frac{r_-^2}{2L^2}}\!\!\left(\!\!
\frac{\cos(k_L r_-)}{\left(\!k_L r_-\!\right)^2}\!-\!\frac{\sin\left(\!k_l r_-\!\right)}{\left(\!k_L r_-\!\right)^3}\!\right)\\
\!\!\!&&\!\!\!\left(\frac{\sin(k_L r_-)}{k_L r_-}+\frac{3\cos(k_L
r_-)}{\left(k_L r_-\right)^2}-\frac{3\sin(k_L r_-)}{\left(k_L
r_-\right)^3}\right),
\end{eqnarray*}
is negligible compared to the first part $\Gamma_{ij,A}$ in the
asymptotic limit $k_L L\rightarrow\infty$. Its integrand tends to
zero for $r_- \rightarrow 0$ (expansion for $r_- \ll 1 $ yields
$k_L^2 r_-^4/45+\mathcal{O}(r_-^5)$) and contains only terms
proportional to $\cos(k_L r_-)\sin(k_L r_-)/(k_L r_-)^{x}$,
$\cos(k_L r_-)^2/(k_L r_-)^{x}$ and $\sin(k_L r_-)^2/(k_L
r_-)^{x}$ with $x\geq 1$. These types of terms have been neglected
in Eq.~(\ref{GammaA}) or decay even faster in $r_-$.
Since $\Gamma_{ij,B}$ is negligible compared to $\Gamma_{ij,A}$,
we use $\Gamma_{ij}=\Gamma\frac{3}{4\left(k_L L\right)^2}$.\\
\\Next, we calculate the imaginary part $G_{ij}=Im\left(\langle
J_{ij}\rangle\right)$ by inserting Eq.~(\ref{ImaginaryPart}) into
Eq.~(\ref{AveragedRate}). As before, we consider the two
contributions $G_{ij}=G_{ij,A}+G_{ij,B}$ corresponding to the
first and the second line in Eq.~(\ref{ImaginaryPart}) separately.
The integrand of the first part
\begin{eqnarray*}
G_{ij,A}&=&-\frac{3 \Gamma}{\sqrt{2 \pi}L^3}\int_{0}^\infty\!\!\!
dr_- r_-^2 e^{-\frac{r_-^2}{2L^2}}
\frac{\cos(k_L r_-)}{k_L r_-}\\
&&\left(\frac{\sin(k_L r_-)}{k_L r_-}+\frac{\cos(k_L
r_-)}{\left(k_L r_-\right)^2}-\frac{\sin(k_L r_-)}{\left(k_L
r_-\right)^3}\right),
\end{eqnarray*}
tends to zero for $r_- \rightarrow 0$ (expansion for $r_-\ll 1$
yields $2 r_-/(3 k_L)+\mathcal{O}(r_-^3)$) and features a rapidly
oscillating term proportional to $\sin\left(k_L
r_-\right)\cos\left(k_L r_-\right)$ in the integral, which leads
to a contribution which scales with $1/(k_L L)^3$. The other terms
proportional to $\cos^2(k_L L)/(k_L L)^3$ and $\cos(k_L L)\sin(k_L
L)/(k_L L)^3$ are again of the type discussed and neglected
before. Hence it is well justified to assume that $G_{ij,I}\ll
\Gamma_{ij,I}$. The integrand of the second part
\begin{eqnarray*}
G_{ij,B}\!\!\!&=&\!\!\!\frac{3 \Gamma}{\sqrt{2
\pi}L^3}\!\!\int_{0}^\infty\!\!\!\!\!dr_-r_-^2
e^{-\frac{r_-^2}{2L^2}}\!\left(\!
\frac{\sin k_L r_-)}{\left(\!k_L r_-\!\right)^2}\!+\!\frac{\cos\left(\!k_L r_-\!\right)}{\left(\!k_L r_-\!\right)^3}\!\right)\\
\!\!\!&&\!\!\!\left(\frac{\sin(k_L r_-)}{k_L r_-}+\frac{3\cos(k_L
r_-)}{\left(k_L r_-\right)^2}-\frac{3\sin(k_L r_-)}{\left(k_L
r_-\right)^3}\right),
\end{eqnarray*}
also tends to zero for $r_- \rightarrow 0$ (expansion for $r_-\ll
1$ yields $-r_-/(15 k_L)+\mathcal{O}(r_-^3)$) and contains only
one term which has not been considered so far. The term in the
integrand proportional $\sin(k_L r_-)^2/(k_L r_-)$ leads to to a
contribution which decays with $\log(k_L L)/(k_L L)^3$ in the
asymptotic limit $k_L L\rightarrow \infty$. The imaginary part of
the averaged decay rate $\langle J_{ij}\rangle$ is therefore
negligible compared to
the real part.\\
\\The distance between the two atomic samples does not
play a role in the setting under consideration. In the limit
$k_{L}\gg R/L^2$, averaged single ensemble rates equal averaged
inter-ensemble rates $\langle J_{ij}\rangle=\langle
J_{ij}\rangle^{I,I}=\langle J_{ij}\rangle^{I,II}$. In the
following, we outline the calculation of the inter-ensemble value
$\Gamma^{I,II}_{ij,A}$. Analogous arguments can be used compute
$\Gamma^{I,II}_{ij,B}$, $G^{I,II}_{ij,A}$, and
$G^{I,II}_{ij,B}$.\\
\\Inter-ensemble rates are obtained by averaging atomic positions
with respect to Gaussian distributions centered at the origin and
a distance $R$ apart respectively
\begin{eqnarray*}
\Gamma_{ij}^{I,II}\!\!\!&=&\!\!\!\frac{1}{\pi^3L^6}\!\int
\!\!d\mathbf{r}\!\!\int\!\! d\mathbf{\acute{r}}
e^{i\mathbf{k}_L\left(\mathbf{r}-\mathbf{\acute{r}}\right)}e^{\frac{-r^2}{L^2}}e^{\frac{-\left(\acute{\mathbf{r}}-\mathbf{R}\right)^2}{L^2}}
\gamma\!\left(\mathbf{r}-\mathbf{\acute{r}}\right)\!,\\
\!\!\!&=&\!\!\!\frac{1}{\left(2\pi\right)^3\!\!L^6}\!\!\!\int
\!\!\!d\mathbf{r_+}\!\!\!\int\!\! \!d\mathbf{r_-}e^{i\mathbf{k}_L
\mathbf{r}_-}e^{\frac{-\left(\mathbf{r}_-+\mathbf{R}\right)^2}{2
L^2}}e^{\frac{-\left(\mathbf{r}_+ -\mathbf{R}\right)^2}{2
L^2}}\!\gamma(\mathbf{r}_-),\\
\!\!\!&=&\!\!\!\frac{1}{\left(2\pi\right)^{3/2}L^3}\int
d\mathbf{r}_-e^{i\mathbf{k}_L
\mathbf{r}_-}e^{\frac{-\left(\mathbf{r}_-+\mathbf{R}\right)^2}{2
L^2}}\gamma(\mathbf{r}_-),
\end{eqnarray*}
where the variable transformation
$\mathbf{r_{+}}=\mathbf{r}+\mathbf{\acute{r}}$,
$\mathbf{r_{-}}=\mathbf{r}-\mathbf{\acute{r}}$ was made, as
before. By inserting Eq.~(\ref{RealPart}) and neglecting the
dipole factor
$\left(\hat{\mathbf{x}}\cdot(\mathbf{r}-\mathbf{\acute{r}})/|\mathbf{r}-\mathbf{\acute{r}}|\right)^2$,
which does not play a role for the distance under consideration
\cite{DipoleFactor}, one obtains
\begin{eqnarray*}
\Gamma_{ij}^{I,II}&=&\frac{3\Gamma}{2\left(2\pi\right)^{3/2}L^3}\int
d\mathbf{ r}_-\ e^{i\mathbf{k}_L
\mathbf{r}_-}e^{\frac{-\left(\mathbf{r}_-+\mathbf{R}\right)^2}{2
L^2}}\frac{\sin\left(k_L r_-\right)}{k_L r_-},\\
&=&\frac{3\Gamma}{2\sqrt{2\pi} L^3 } \int_{0}^{\infty} \!\!\!d
r_-r_-^2\frac{\sin\left(k_L
r_-\right)}{k_Lr_-}\!\!\!\int_0^{\pi}\!\!\! d\theta\sin(\theta)\\
&&e^{i k_L r_-\cos(\theta)}
e^{\frac{-1}{2L^2}\left(r_-^2+R^2+2r_-R\cos(\theta)\right)},\\
&=&\frac{3\Gamma}{2\sqrt{2\pi} L^3 k_L}\int_{0}^{\infty} \!\!\!d
r_-\sin\left(k_L
r_-\right)e^{\frac{-\left(r_-^2+R^2\right)}{2L^2}}\\
&&\int_{-r_-}^{r_-}dx e^{-i k_L x}e^{\frac{Rx}{L^2}}.
\end{eqnarray*}
In the last step, the integral over $\theta$ was transformed using
the variable transformation $x=-\cos(\theta)r_-$. The integral
over $x$ can be directly evaluated yielding
\begin{eqnarray*}
\Gamma_{ij}^{I,II}&=&\frac{3\Gamma}{2\sqrt{2\pi} L^3 k_L\left(i
k_L -R/L^2\right)}\int_{0}^{\infty} \!\!\!d r_-\sin\left(k_L
r_-\right)\\
&&e^{\frac{-\left(r_-^2+R^2\right)}{2L^2}} \left(e^{ik_L
r_--\frac{R r_-}{L^2}}-e^{-ik_L r_-+\frac{R r_-}{L^2}}\right),\\
&=&\frac{3\Gamma}{2\sqrt{2\pi} L^3 k_L\left(i k_L
-R/L^2\right)}\int_{-\infty}^{\infty} \!\!\!d r_-\sin\left(k_L
r_-\right)\\
&&e^{\frac{-\left(r_-+R\right)^2}{2L^2}}e^{i k_L r_-}.
\end{eqnarray*}
As next step, the variable transformation $\tilde{r}=r_-+R$ is
made such that
\begin{eqnarray*}
\Gamma_{ij}^{I,II}&=&\frac{3\Gamma}{2\sqrt{2\pi} L^3 k_L\left(i
k_L -R/L^2\right)}\int_{-\infty}^{\infty} \!\!\!d
\tilde{r}\sin\left(k_L
(\tilde{r}-R)\right),\\
&&e^{\frac{-\tilde{r}^2}{2L^2}}e^{i k_L (\tilde{r}-R)}\\
&=&\frac{3\Gamma}{4\sqrt{2\pi} L^3 k_L\left(-k_L
-iR/L^2\right)}\int_{-\infty}^{\infty} \!\!\!d
\tilde{r}e^{\frac{-\tilde{r}^2}{2L^2}},\\
&&(e^{2ik_L(\tilde{r}-R)}-1),\\
&=&\frac{3\Gamma}{4 \left(L^2 k_L^2+ik_L R\right)}\left(1-e^{-2
k_L^2 L^2-2ik_L R}\right)
\end{eqnarray*}
which yields $\Gamma_{ij}^{I,II}=\Gamma\frac{3}{4\left(k_L
L\right)^2}$ for $k_{L}\gg R/L^2$ and $k_L L\gg1$.
\section{Time evolution of entanglement in a two-level model}\label{Appendix1}
In this appendix, we calculate the amount of entanglement produced
$\xi(t)$ (compare Eq.~(\ref{xi})) for the model described in
Sec.~\ref{TwoLevelModel}. The first part of this appendix
\ref{Ent} covers the derivation of $\xi(t)$ based on the full
master equation (\ref{ME3}). The second part \ref{FullME} contains
explanations concerning Eq.~(\ref{ME3}).
\subsection{Time evolution of entanglement}\label{Ent}
In the following, the time evolution of entanglement $\xi(t)$ is
calculated. To this end we calculate the single-ensemble variance
of transverse spin components $\text{var}(J_{z})=\langle
J_z^2\rangle-\langle J_z\rangle^2$, the inter-ensemble product of
transverse spins $\langle J_{z,I} J_{z,II}\rangle$ and finally the
mean value of the longitudinal spin $\langle J_x \rangle$.\\
\\We start by calculating  $d_t\langle J_z^2 \rangle_t$. The
dissipative evolution described by Eq.~(\ref{ME3}) leads to
\begin{eqnarray*}
d_t\langle J_z^2\rangle\!\!&=&\!\!\!- \tilde{\Gamma}\langle
J_z^2\rangle-\frac{2d}{N}\Gamma \langle J_z^2
J_x\rangle+\!\frac{N}{4}\tilde{\Gamma} +\!\frac{d\Gamma}{N}
\langle J_x^2\rangle\!\left(\mu^2+\nu^2\right)\!.
\end{eqnarray*}
Applying the decorrelation approximation $\langle J_z
J_x\rangle\approx\langle J_z \rangle\langle J_x \rangle$
\cite{DecorrelationApproximation}
for mean values of products of transverse and longitudinal spin
yields
\begin{eqnarray*}
d_t\langle J_z^2\rangle\!\!\!&=&\!\!\!-
\left(\tilde{\Gamma}\!+\!d\Gamma P_2(t)\!\right)\!\langle
J_z^2\rangle\!+\!\frac{N}{4}\!\left(\tilde{\Gamma}+d\Gamma
P_2(t)^2\!\!\left(\mu^2+\nu^2\right)\!\right)
\end{eqnarray*}
and similarly
\begin{eqnarray*}
d_t\langle J_y^2\rangle\!\!\!&=&\!\!\!-
\left(\tilde{\Gamma}\!+\!d\Gamma P_2(t)\!\right)\!\langle
J_y^2\rangle\!+\!\frac{N}{4}\!\left(\tilde{\Gamma}+d\Gamma
P_2(t)^2\!\!\left(\mu^2+\nu^2\right)\!\right)
\end{eqnarray*}
where $\langle J_x\rangle=\frac{N}{2}P_2(t)$ and $\langle
J_x^2\rangle\approx \langle J_x\rangle^2=\frac{N^2}{4}P_2(t)^2$
were used.
The latter approximation leads only to an error of the order
$\mathcal{O}\left(\frac{1}{N}\right)$ \cite{Error}
Next, the mean values of the transverse spin components are
computed using the same approximations.
\begin{eqnarray*}
d_{t}\langle
J_{y/z}\rangle_t&=&-\frac{1}{2}\left(\tilde{\Gamma}+d\Gamma
P_2(t)\right)\langle J_{y/z}\rangle_t,
\end{eqnarray*}
with $\langle J_{y/z}\rangle_{t=0}=0$. The mean values can
therefore be ignored when calculating single ensemble
variances $\text{var}(J_{y/z})$.\\
\\The time derivatives of inter-ensemble products of transverse
spins are given by
\begin{eqnarray*}
d_t\langle J_{z,I} J_{z,II}\rangle\! \!=\!
\!-\!\left(\tilde{\Gamma}+d\Gamma P_2(t)\right)\langle J_{z,I} J_{z,II}\rangle\!+\!\frac{N}{2}\mu\nu d\Gamma P_2(t)^2,\\
d_t\langle J_{y,I} J_{y,II}\rangle\! \!=\!
\!-\!\left(\tilde{\Gamma}+d\Gamma P_2(t)\right)\langle J_{y,I}
J_{y,II}\rangle\!-\!\frac{N}{2}\mu\nu d\Gamma P_2(t)^2,
\end{eqnarray*}
where we used $\langle J_{x,I} J_{x,II}\rangle\approx\langle
J_{x,I}\rangle \langle J_{x,II}\rangle$. For $N\gg 1$, this is a
very good approximation, since collective effects on populations
are suppressed by a factor $d/N$. Therefore, the time evolution of
longitudinal spins is only determined by single-particle terms,
which do not lead to correlations between the two ensembles.
Hence, the variances of the non-local operators
$J_{y,\pm}=\left(J_{y,I}\pm J_{y,II}\right)/\sqrt{2}$ and
$J_{z,\pm}=\left(J_{z,I}\pm J_{z,II}\right)/\sqrt{2}$ evolve
according to
\begin{eqnarray}\label{2Derivative}
d_{t}\text{var}\left(J_{y,\pm}\right)&=&-\left(\tilde{\Gamma}+d\Gamma
P_2(t)\right)\text{var}\left(J_{y,\pm}\right)\\
&+&\frac{N}{4}\left(\tilde{\Gamma}+d\Gamma
P_2(t)^2\left(\mu\mp\nu\right)^2\right),\nonumber\\
d_{t}\text{var}\left(J_{z,\pm}\right)&=&-\left(\tilde{\Gamma}+d\Gamma
P_2(t)\right)\text{var}\left(J_{z,\pm}\right)\\
&+&\frac{N}{4}\left(\tilde{\Gamma}+d\Gamma
P_2(t)^2\left(\mu\pm\nu\right)^2\right),\nonumber
\end{eqnarray}
such that
\begin{eqnarray*}
\text{var}\left(J_{y,\pm}\right)_{\infty}=\text{var}\left(J_{z,\mp}\right)_{\infty}=\frac{N}{4}\frac{\tilde{\Gamma}+d\Gamma
P_{2,\infty}^2\left(\mu\mp\nu\right)^2}{\tilde{\Gamma}+d\Gamma
P_{2,\infty}}
\end{eqnarray*}
in the steady state. The variances
$\text{var}\left(J_{y,+}\right)$ and
$\text{var}\left(J_{z,-}\right)$ are squeezed, while
$\text{var}\left(J_{y,-}\right)$ and
$\text{var}\left(J_{z,+}\right)$ are anti-squeezed. Now, we
consider the time evolution of the longitudinal spin
\begin{eqnarray*}
d_{t}|\langle
J_x\rangle|&=&-\frac{N}{2}\left({\Gamma}_{\text{heat}}-{\Gamma}_\text{cool}\right)-\left({\Gamma}_\text{heat}+{\Gamma}_\text{cool}\right)|\langle
J_x\rangle|,
\end{eqnarray*}
which yields directly
\begin{eqnarray*}
|\langle J_x\rangle|_\infty=\frac{N}{2}\
\frac{{\Gamma}_{\text{cool}}-{\Gamma}_{\text{heat}}}{{\Gamma}_{\text{cool}}+{\Gamma}_{\text{heat}}}=\frac{N}{2}P_{2,\infty}
\end{eqnarray*}
for $t\rightarrow\infty$.\\
\\Collective effects have an negligible effect on the time
evolution of the polarization. $P_2(t)$ evolves due to
single-particle effects only and hence much slower than
$\text{var}\left(J_{y,\pm}\right)$ and
$\text{var}\left(J_{z,\pm}\right)$ for samples with high optical
depth. In this case, the solution for
$\xi(t)=\text{var}\left(J_{z,+}\right)/P_2(t)$ can be cast in a
simple analytical form
\begin{eqnarray*}
\xi(t)\!\!&=&\!\!\frac{1}{P_2(t)}e^{-\left(\tilde{\Gamma}+d\Gamma
P_2(t)\right)t}\\
\!\!&+&\!\!\frac{1}{P_2(t)}\frac{\tilde{\Gamma}+d\Gamma
P_2(t)^2(|\mu|-|\nu|)^2}{\tilde{\Gamma}+d\Gamma
P_2(t)}\left(1-e^{-\left(\tilde{\Gamma}+d\Gamma
P_2(t)\right)t}\right).
\end{eqnarray*}
\subsection{Full master equation}\label{FullME}
In the following, we comment on the form of Eq.~(\ref{ME3}),
in particular on the absence of collective noise terms.\\
\\The probe fields considered in Sec.~\ref{TwoLevelModel} are
off-resonant and it has been shown in the main text that
collective contributions feature an enhancement factor which
renders them the dominant decay mechanism for samples with high
optical depth. As is shown in Sec.~\ref{CreationOfEntanglement},
it can be advantageous to apply also resonant laser light (pump
fields). In contrast to off-resonant fields, collective
contributions are negligible compared to single-particle terms for
resonant light in the situation considered here. Unlike
off-resonant collective rates, resonant collective rates are much
slower than the corresponding single-particle rates for samples
with high optical depth, which is an effect well known and
harnessed in electromagnetically induced transparency
\cite{EIT,EITreview,AlexeyI,Jonatan}. The single particle decay
rate after adiabatic elimination of excited states is given by
$\Gamma_{res}=\frac{\Omega_{\text{pump}}^2}{\gamma_{\text{LW}}}$,
where $\Omega_{\text{pump}}$ is the Rabi frequency of the applied
laser field and $\gamma_{\text{LW}}$ is the natural line width of
excited levels. Coherent effective effects lead to an enhancement
factor $d$ in the denominator. Intuitively, this effect can be
understood by noting that emitted resonant photons
are reabsorbed in an optically thick medium.\\
\\Due to atomic motion in ensembles at room temperature, spectral
lines are Doppler broadened. We take therefore off-resonant
contributions of pump fields to the master equation with a
detuning on the order of  the Doppler width
$\delta_{\text{Doppler}}$ into account. A calculation along the
lines of the derivation shown in Sec.~\ref{Ent} shows that these
terms are negligible compared to their single-atom counterparts.
More specifically, collective terms corresponding to a detuning
$\delta_{\text{Doppler}}$ lead to decay rates proportional to
$\frac{\Omega_{\text{pump}}^2}{\delta_{\text{Doppler}}}
\gamma_{\text{LW}}
d=\frac{\Omega_{\text{pump}}^2}{\gamma_{\text{LW}}
d}\left(\frac{\gamma_{\text{LW}}
d}{\delta_{\text{Doppler}}}\right)^2$ with
$|\delta_{\text{Doppler}}|>>\gamma_{\text{LW}} d$, while single
particle resonant terms lead to decay rates proportional to
$\frac{\Omega_{\text{pump}}^2}{\gamma_{\text{LW}}}$.\\
\\Finally, Eq.~(\ref{ME3}) does not include collective terms,
corresponding to radiative processes which do not change the
internal atomic state, since they do not have an effect on the
amount of entanglement generated.
A master equation corresponding to the terms omitted in Eq.~
(\ref{ME2})
\begin{eqnarray*}
d_t\rho(t)&=&d\frac{\check{\Gamma}}{2} C\rho(t)
C^{\dag}+d\frac{\check{\Gamma}}{2} D\rho(t) D^{\dag}+...\ ,
\end{eqnarray*}
with operators $C=\sum_i\left(\mu\
\sigma_{\downarrow\downarrow,I,i}+\nu\
\sigma_{\uparrow\uparrow,II,i}\right)$ and $D=C=\sum_i\left(\mu\
\sigma_{\downarrow\downarrow,II,i}+\nu\
\sigma_{\uparrow\uparrow,I,i}\right)$ leads to $d_t\xi(t)=0$.
Since $d_t\langle J_y\rangle=-\frac{d \check{\Gamma}}{2 N}\langle
J_y\rangle$, $d_t\langle J_z\rangle=-\frac{d \check{\Gamma}}{2
N}\langle J_z\rangle$, and $d_t\langle J_x\rangle=0$, $\langle
J_y\rangle=\langle J_z\rangle=0$ and $\langle J_z\rangle=N/2$ for
all times. The time derivatives of single-ensemble variances for
transverse spin components is given by
\begin{eqnarray*}
d_t \langle J_y^2\rangle
&=&\check{\Gamma}\frac{d}{N}\left(\mu^2+\nu^2\right)\left(-\langle
J_y^2\rangle+\langle J_z^2\rangle\right),\\
d_t \langle J_z^2\rangle
&=&\check{\Gamma}\frac{d}{N}\left(\mu^2+\nu^2\right)\left(\langle
J_y^2\rangle-\langle J_z^2\rangle\right),
\end{eqnarray*}
such that $\langle J_y^2\rangle=\langle J_z^2\rangle=N/4$, for all
times. Accordingly, $\langle J_{y,I}J_{y,II}\rangle=\langle
J_{z,I}J_{z,II}\rangle=0$ for all times since
\begin{eqnarray*}
d_t\langle
J_{y,I}J_{y,II}\rangle &=&-2\frac{d}{N}\hat{\Gamma}\mu\nu\langle J_{z,I}J_{z,II}\rangle,\\
d_t\langle J_{z,I}J_{z,II}\rangle
&=&2\frac{d}{N}\hat{\Gamma}\mu\nu\langle J_{y,I}J_{y,II}\rangle.
\end{eqnarray*}
The processes under consideration do not create entanglement
unlike the terms in Eq.~(\ref{ME2}) with jump operators $A$ and
$B$. As shown above, they do not degrade entanglement either.
Collective terms corresponding to far off-resonant radiative
transitions
$|\uparrow\rangle\rightarrow|e_{\downarrow}\rangle\rightarrow|\uparrow\rangle$,
$|\downarrow\rangle\rightarrow|e_{\uparrow}\rangle\rightarrow|\downarrow\rangle$
do not introduce random phases and preserve coherence. The emitted
photon does not reveal information about the internal atomic
state, since it is emitted into the laser mode. Terms with jump
operators $C$ and $D$ lead only to very small correction terms
proportional to $1/N$ and can be ignored.
\section{Generation of steady state entanglement in alkali atoms}\label{ThreeLevelCalculation}
In this appendix, we consider the generation of dissipatively
driven entanglement in multi-level systems based on the model
described in Sec.~\ref{MultilevelDynamics}.\\
\\Taking three ground state levels $|\!\!\uparrow\rangle$,
$|\!\!\downarrow\rangle$ and $|h\rangle$ into account, as shown in
Fig.~\ref{Alkalilevels}, the evolution of the reduced atomic
density matrix can be described by the master equation
\begin{eqnarray}\label{ME4}
d_{t}\rho(t)\!\!&=&\!\! d\Gamma  A \rho(t)
A^{\dag} + d\Gamma  B \rho(t) B^{\dag}\\
\!\!\!\!\!\!&+&\!\!\Gamma_{\downarrow\!\uparrow}\!\sum_{i=1}^{N}\left(
\sigma_{I,i}\rho(t)\sigma_{I,i}^{\dag}+
\sigma_{II,i}\rho(t)\sigma_{II,i}^{\dag}\right)\nonumber\\
&+&\!\!\Gamma_{\uparrow\!\downarrow}\!\sum_{i=1}^{N}\left(
\sigma_{I,i}^{\dag}\rho(t)\sigma_{I,i}+
\sigma_{II,i}^{\dag}\rho(t)\sigma_{II,i}\right)\nonumber\\
\!\!&+&\!\!\Gamma_{\uparrow \!
h}\!\sum_{i=1}^{N}\left(|h\rangle\!_{I,i}\langle\uparrow\!\!|\rho(t)
|\!\!\uparrow\rangle\!_{I,i}\langle h|\! +\!
|h\rangle\!_{II,i}\langle\uparrow\!\!|\rho(t)
|\!\!\uparrow\rangle\!_{II,i}\langle h|\right)\nonumber\\
\!\!&+&\!\!\Gamma_{h\!
\uparrow}\!\sum_{i=1}^{N}\left(|\!\!\uparrow\rangle\!_{I,i}\langle
h|\rho(t) |h\rangle\!_{I,i}\langle\!\uparrow\!\!| +
|\!\!\uparrow\!\rangle\!_{II,i}\langle h|\rho(t)
|\!h\rangle\!_{II,i}\langle \!\uparrow\!\!|\right)\nonumber\\
\!\!&+&\!\!\Gamma_{\downarrow\!
h}\!\sum_{i=1}^{N}\left(|\!h\rangle\!_{I,i}\langle\downarrow\!\!|\rho(t)
|\!\!\downarrow\rangle\!_{I,i}\langle h| +
|\!h\rangle\!_{II,i}\langle\downarrow\!\!|\rho(t)
|\!\!\downarrow\rangle\!_{II,i}\langle h|\right)\nonumber\\
\!\!&+&\!\!\Gamma_{h\!
\downarrow}\!\sum_{i=1}^{N}\left(|\!\!\downarrow\rangle\!_{I,i}\langle
h|\rho(t) |h\rangle\!_{I,i}\langle \!\downarrow\!\!| +
|\!\!\downarrow\!\rangle\!_{II,i}\langle h|\rho(t)
|h\rangle\!_{II,i}\langle \!\downarrow\!\!|\right)\nonumber\\
\!\!&+&\!\!\Gamma_{\uparrow\!\uparrow}\!\sum_{i=1}^{N}\left(|\!\!\uparrow\rangle\!_{I,i}\langle\uparrow\!\!|\rho(t)
|\!\!\uparrow\rangle\!_{I,i}\langle\uparrow\!\!| +
|\!\!\uparrow\rangle\!_{II,i}\langle\uparrow\!\!|\rho(t)
|\!\!\uparrow\rangle\!_{II,i}\langle\uparrow\!\!|\right)\nonumber\\
\!\!&+&\!\!\Gamma_{\downarrow\!\downarrow}\!\sum_{i=1}^{N}\left(|\!\!\downarrow\rangle\!_{I,i}\langle\downarrow\!\!|\rho(t)
|\!\!\downarrow\rangle\!_{I,i}\langle\downarrow\!\!| +
|\!\!\downarrow\rangle\!_{II,i}\langle\downarrow\!\!|\rho(t)
|\!\!\downarrow\rangle\!_{II,i}\langle\downarrow\!\!|\right),\nonumber
\end{eqnarray}
where $\Gamma_{ab}$ is the single-particle rate for the transition
$|a\rangle\rightarrow|b\rangle$.
As in Sec.~\ref{MasterEquation2}, we omit collective terms due to
resonant pump fields, as well as collective dephasing terms (see
App.~\ref{FullME}). Collective terms involving the level
$|h\rangle$ are also insignificant for $N>>1$, as long as the
number of coherent collective excitations is small.\\
\\In order to compute the amount of entanglement produced,
we consider the variance of the nonlocal operator
$J_{y,+,2}=\left(J_{y,I}+J_{y,II}\right)_2/\sqrt{2}$. (The
subscript "2" emphasizes that these quantities are defined with
respect to the two-level subsystem
$\{|\!\!\uparrow\rangle,|\!\!\downarrow\rangle\}$.) A calculation
analogous to the two-level derivation in App.~\ref{Ent} shows that
$\langle J_{y,I}\rangle_2=\langle J_{y,II}\rangle_2=0$ for all
times. Therefore $\text{var}\left(J_y\right)_2=\langle
J_y^2\rangle_2$. For simplicity, we assume that both ensembles are
identical $\langle J_{y,I}^2\rangle_2=\langle
J_{y,II}^2\rangle_2=\langle
J_{y}^2\rangle_2$.\\
According to Eq.~(\ref{ME4}), the time derivative of the
single-ensemble variance $\langle J_{y}^2\rangle_2$ is given by
\begin{eqnarray*}
d_t \langle J_{y}^2\rangle_2&=&-\left(\bar{\Gamma}+d \Gamma
\frac{N_2(t)}{N}P_2(t)\right)\langle J_{y}^2\rangle_2+\Gamma
\frac{N_2(t)}{4}\\
&+&d\Gamma\frac{1}{4}\frac{N_2(t)}{N}P_2(t)^2\left(\mu^2+\nu^2\right),
\end{eqnarray*}
where the decay rate $\bar{\Gamma}$, the number of atoms in the
relevant two-level subsystem
$\{|\!\!\uparrow\rangle,|\!\!\downarrow\rangle\}$, $N_2(t)$ and
the corresponding polarization $P_2(t)$ are defined in
Sec.~\ref{MultilevelCalculation}.
$N=\sum_i\left(|\!\!\uparrow\rangle_i\langle
\uparrow|+|\!\!\downarrow\rangle_i\langle
\downarrow|+|h\rangle_i\langle h|\right)$ is the total number of
atoms in one ensemble and $N_2(0)=N$.
Note that repump fields, which transfer atoms from $|h\rangle$ to
$|\!\!\uparrow\rangle$ or $|\!\!\downarrow\rangle$ (corresponding
to terms with prefactors $\Gamma_{hg}$ and $\Gamma_{hs}$ in
Eq.~(\ref{ME4})) do not contribute to $\bar{\Gamma}$.\\
\\Inter-ensemble correlations $\langle J_{y,I}J_{y,II}\rangle_2$
evolve according to
\begin{eqnarray*}
d_t \langle J_{y,I}J_{y,II}\rangle_2&=&-\left(\bar{\Gamma}+d
\Gamma \frac{N_2(t)}{N}P_2(t)\right)\langle
J_{y,I}J_{y,II}\rangle_2\\
&-&d\Gamma \frac{1}{2}\frac{N_2(t)}{N}P_2(t)^2 \mu\nu.
\end{eqnarray*}
Hence, the time evolution of
$\text{var}\left(J_{y,+}\right)_2=\langle J_{y}^2\rangle_2+\langle
J_{y,I}J_{y,II}\rangle_2$ is given by
\begin{eqnarray}\label{3Derivative}
d_t \langle J_{y,+}^2\rangle_2&=&-\left(\bar{\Gamma}+d \Gamma
\frac{N_2(t)}{N}P_2(t)\right)\langle J_{y,+}^2\rangle_2+\bar{\Gamma}\frac{N_2(t)}{4}\nonumber\\
&+&d\Gamma \frac{1}{4}\frac{N_2(t)}{N}P_2(t)^2
\left(\mu-\nu\right)^2.
\end{eqnarray}
Analogously,
\begin{eqnarray}\label{3Derivative}
d_t \langle J_{z,-}^2\rangle_2&=&-\left(\bar{\Gamma}+d \Gamma
\frac{N_2(t)}{N}P_2(t)\right)\langle J_{z,-}^2\rangle_2+\bar{\Gamma}\frac{N_2(t)}{4}\nonumber\\
&+&d\Gamma \frac{1}{4}\frac{N_2(t)}{N}P_2(t)^2
\left(\mu-\nu\right)^2.
\end{eqnarray}
Since the evolution of $N_2(t)$ and $P_2(t)$ is known from
equations (\ref{Rate equations}), $\Sigma_{J,2}=\langle
J_{y,+}^2\rangle_2+\langle J_{z,-}^2\rangle_2$ can be directly
calculated yielding a complicated expression. However, as
explained in Sec.~\ref{TwoLevelModel} and Sec.~\ref{Alkali
Implementation}, $N_2(t)$ and $P_2(t)$ can be considered to change
slowly compared to the fast entangling dynamics. In this case,
(\ref{3Derivative}) leads to the simple and convenient expression
(\ref{3Ent}) used in Sec.~\ref{MultilevelCalculation}.
\section{Implementation in hot $133\text{Cs}$ vapors}\label{Appendix2}
In this appendix, we apply the results derived in Sec.~\ref{Alkali
Implementation} to a specific example and consider the generation
of entanglement between two $^{133}\text{Cs}$ ensembles at room
temperature. The parameters used in the following take values
consistent with the experiments reported in \cite{Polzik2001,
Squeezing}. The approximate calculation outlined below provides a
rough estimate of the entanglement that can be produced.
\\We assume $\hat{\mathbf{y}}$-polarized probe light which propagates along
$\hat{\mathbf{z}}$ and interacts in succession with two ensembles
in a magnetic field which is oriented along $\hat{\mathbf{x}}$.
The laser field is assumed to be blue detuned by $\Delta=700
\text{MHz}$ with respect to the $6S_{1/2}(F=4)\rightarrow
6P_{3/2}(F=5)$ transition (D2 line). Fig.~\ref{Cesiumlevels}
depicts the relevant parts of the atomic level schemes in both
samples and illustrates the atomic transitions due to the
light-matter interaction induced by the applied laser field.
Initially, all atoms are pumped to state $|\!\!\uparrow\rangle$.
The restriction of the analysis to the three levels
$|\!\!\uparrow\rangle$, $|\!\!\downarrow\rangle$ and $|h\rangle$
in the presence of strong pump fields, as described in
Sec.~\ref{MultilevelDynamics} for $\hat{\mathbf{x}}$-polarized
probe light is also valid for this configuration, as the rates of
transitions from level $|\!\!\uparrow\rangle$ to states with
$m_{F}=\pm 2$ occur at rates which are two orders of magnitude
smaller than transitions within the sub-system under
consideration. ($\Gamma_{|4,4\rangle\rightarrow|4,2\rangle}=0.03\
\Gamma_{|4,4\rangle\rightarrow|4,3\rangle}$ and
$\Gamma_{|4,4\rangle\rightarrow|3,2\rangle}=0.02\
\Gamma_{|4,4\rangle\rightarrow|4,3\rangle}$).\\
\\In order to calculate the experimentally measurable steady state
entanglement using Eq.~(\ref{ExpEnt}) for a given optical depth
$d$ and parameters $\mu$ and $\nu$, $N_2(t)$, $P_2(t)$ and
$\bar{\Gamma}$ need to be computed.
$N_2(t)=N_{\uparrow}(t)+N_{\downarrow}(t)$ and
$P_2(t)=\left(N_{\uparrow}(t)-N_{\downarrow}(t)\right)/N_2(t)$ are
readily obtained from Eq.~(\ref{Rate equations}), if the rates for
all transitions are known.
As the probe field is assumed to be off-resonant, probe induced
rates $\Gamma_{\text{ab}}$ for transitions
$|a\rangle\rightarrow|b\rangle$ are calculated using the formula
$\Gamma^{\text{probe}}_{ab}=\Omega^2_{\text{probe}}\sum_{l}\frac{(c^{\text{ab}}_l)^2}{\Delta^{\text{ab}}_{l}}\gamma_{\text{LW}}$,
where the sum runs over all levels contributing to a particular
transitions (for example the states $|5,3\rangle$, $|4,3\rangle$
and $|3,3\rangle$ in $6^2P_{3/2}$ if
$\Gamma_{\uparrow\downarrow}=\Gamma \nu^2$ is computed).
$\Delta^{\text{ab}}_l$ is the detuning for each contributing
level, $\gamma_{\text{LW}}$ the natural line width of excited
levels and $c^{\text{ab}}_l$ is the product of the corresponding
Clebsch Gordan coefficients.
Pump, or repump induced transitions are resonant and involve to a
good approximation only one level. Hence the corresponding rates
are given by
$\Gamma_{\text{ab}}^{\text{pump}}=\frac{\Omega^2_{\text{pump}}}
{\gamma_{\text{LW}}}c_{\text{pump}}^2 \ k$ and
$\Gamma_{\text{ab}}^{\text{repump}}=\frac{\Omega^2_{\text{repump}}}
{\gamma_{\text{LW}}}c_\text{repump}^2\ k$ respectively, where
$k=\frac{\Gamma}{\delta{\text{Doppler}}}=\frac{5\text{MHz}}{380\text{MHz}}$
\cite{Doppler}. Here, we consider pumping resonant with respect to
the level $|4,4\rangle$ in $6P_{1/2}$ ($\text{D}_1$ line) and pump
fields resonant with respect to the level $|4,4\rangle$ in
$6P_{3/2}$ ($\text{D}_2$ line). (The decay rates are approximately
the same $\gamma_{D_1}\approx \gamma_{D_2}=\gamma_{\text{LW}}$.)
Having expressions for $N_2(t)$, $P_2(t)$ as well as
$\bar{\Gamma}=\Gamma_{\uparrow\!\downarrow}+\Gamma_{\downarrow\!\uparrow}+\Gamma_{\uparrow\!h}+\Gamma_{\downarrow\!h}+\Gamma_{\uparrow\!\uparrow}+\Gamma_{\downarrow\!\downarrow}+r$
at hand, the amount of entanglement which can be produced in this
particular setting can be calculated. Results are shown in
Figs.~\ref{CesiumPlot1} and \ref{CesiumPlot2} and are discussed in
Sec.~\ref{MultilevelCalculation}. As the three-level description
becomes inaccurate if too many atoms are transferred from state
$|\!\!\uparrow\rangle$ to state $|\!\!\downarrow\rangle$, both
plots show results for the optimal (that is minimal) pump power
which guarantees a fraction of at least $95\%$ of all atoms in the
relevant two-level subsystem in state $|\!\!\uparrow\rangle$ for
all times. Besides the need for sufficient pump fields, repumping
of atoms from $F=3$ to $F=4$ is required. If strong pump- but no
repump fields are applied, no entangled state can be reached, as
for $t\rightarrow\infty$, all atoms are transferred to level
$|h\rangle$.
%
%
%

%

\begin{thebibliography}{54}
\bibitem{AB} We use operators $\tilde{A}$, $\tilde{B}$ labelled with
a tilde for bosonic modes. Later in the text, operators $A$, $B$
without a tilde are used to denote linear combinations of spins.
\bibitem{FTspins}The quantization axis is defined by the direction of the
magnetic field, which is oriented along $\hat{\mathbf{x}}$, hence
$j_x=\frac{1}{2}\left(|\!\!\uparrow\rangle\langle
\uparrow\!\!|-|\!\!\downarrow\rangle\langle
\downarrow\!\!|\right)$.
\bibitem{FTapproxmation} For large, strongly polarized atomic ensembles, collective spins can
be described by bosonic modes $J^{-}_I\propto a$,
$J^{-}_{II}\propto b$ using the Holstein-Primakoff approximation.
The results derived here are based on related approximations and
are valid for $N\gg1$ (compare App.~\ref{Appendix1}).
\bibitem{Analogy}The basic working principle of this approach can be illustrated by
considering two single atoms with ground state $|g\rangle$ and
excited state $|e\rangle$, which are placed in two different
cavities. Both atoms are initially prepared in the excited state
$|\Psi^{\text{in}}\rangle=|e\rangle_{1}|e\rangle_{2}$. Atoms can
spontaneously decay to their ground state $|e\rangle\rightarrow
|g\rangle$ while emitting a photon. The light emitted from both
cavities is combined at a beamsplitter and photon detectors are
placed at each output port. The atoms are projected into a
maximally entangled state
$|\Psi^{\text{out}}\rangle=\left(|e\rangle_1|g\rangle_2+|g\rangle_1|e\rangle_2\right)/\sqrt{2}$
if a photon is detected at one of the two output ports.
\bibitem{Normalization} $\mu$ and $\nu$ are normalized such that
$\mu^2-\nu^2=1$ and accordingly, $[A,A^{\dag}]=[B,B^{\dag}]=1$.
\bibitem{HalfWavePlate} A $\lambda/2$ plate interchanges the fields
in $\hat{\mathbf{x}}$ and $\hat{\mathbf{y}}$ polarization. If such
a passive optical element is placed between the first and the
second atomic ensemble, the roles of classical and quantum fields
are interchanged. Therefore, the classical field drives
$\pi$-transitions $|\!\!\uparrow\rangle\rightarrow
|e_{\uparrow}\rangle$, $|\!\!\downarrow\rangle\rightarrow
|e_{\downarrow}\rangle$, in the second ensemble, while the quantum
field is associated with  transitions which change the magnetic
quantum number $\Delta_m=\pm 1$.
\bibitem{UndesiredTransitions} Undesired radiative transitions $|\!\uparrow\rangle\rightarrow |\!\uparrow\rangle$
and $|\!\downarrow\rangle\rightarrow |\!\downarrow\rangle$ involve
the emission of a photon but no change of the internal atomic
state. These processes are not explicitely shown in
Fig.~\ref{Figure2}.
\bibitem{IndependentBaths} If the upper and lower sideband are not treated as independent baths,
cross terms appear, which rotate fast (with frequency $\pm 2
\Omega$) compared to the single-bath terms, (which do not rotate
in this picture). These cross terms can be neglected in a rotating
wave approximation if $\Omega>>\Gamma_{\text{atomic}}$.
\bibitem{Rates} Rates $\Gamma_{ab}$ are calculated using the formula
$\Gamma_{ab}=\Omega_{ab}^2\frac{c_{ab}^2}{|\Delta_{ab}+i\gamma_{\text{LW}}|^2}\gamma_{\text{LW}}
$, where $\Omega_ {ab}$ is the Rabi frequency, $\Delta_ {ab}$ the
detuning, $\gamma_{\text{LW}}$ the natural line width of excited
levels and $c_{ab}$ the Clebsch Gordan coefficient for the
transition under consideration. The cooling rate
$\Gamma_{\text{cool}}=\Gamma^{\text{probe}}_{\text{cool}}+\Gamma^{\text{pump}}_{\text{cool}}$
consists of a probe- and a pump induced part. As probe fields are
considered to be far off-resonant, the corresponding cooling rate
is calculated using the approximation
$\Gamma^{\text{probe}}_{\text{cool}}=\Omega^2_{\text{probe}}\frac{1}{\left(\Delta-\Omega\right)^2}\gamma_{\text{LW}}$.
Contributions due to resonant pump fields are given by
$\Gamma^{\text{pump}}_{\text{cool}}=\Omega^2_{\text{pump}}\frac{1}{\gamma^2_{\text{LW}}}\gamma_{\text{LW}}$.
\bibitem{Doppler} The correction factor $k$ takes into account that due to the Doppler broaddening of atoms moving at room temperature only a fraction
$k=\frac{\gamma_{\text{LW}}}{\delta_{\text{Doppler}}}$ of all
atoms in the cell is on resonance with the applied field. The
Doppler width is given by
$\delta_{\text{Doppler}}=\frac{\nu}{c}\left(\frac{2 k_{B}T}{m \ln
2}\right)^{1/2}$, where $c/\nu=\lambda$ is the wavelength of the
applied light field, $k_B$ is the Boltzmann constant, $T$ is the
temperature and $m$ is the atomic mass.
\bibitem{Vacuum} Eq.~(\ref{ZeroTemperatureBath}) yields $d_{t}\langle
a^{\dag}a\rangle=-\kappa_{A}\langle a^{\dag}a\rangle$ and
$d_{t}\langle b^{\dag}b\rangle=-\kappa_{B}\langle
b^{\dag}b\rangle$. Hence, $\langle a^{\dag}a\rangle_\infty=\langle
b^{\dag}b\rangle_\infty=0$ in the steady state.
\bibitem{DipoleFactor} The dipole term
$\left(\hat{\mathbf{p}}\cdot\hat{\mathbf{r}}_{ij}\right)^2)$ leads
to contributions which decay quickly in the interatomic distance.
The corresponding terms in Eq.~(\ref{GammaA}) (the second and
third term in brackets) are also negligible in the calculation of
the single-ensemble rate $\Gamma_{ij,A}$ as discussed in
App.~\ref{DerivationME-II}.
\bibitem{DecorrelationApproximation} For perfectly polarized ensembles $ \langle J_z J_x\rangle=\langle J_z \rangle\langle J_x \rangle$. The
decorrelation approximation is justified as long as only a small
number of collective (coherent) excitations is created in the
atomic sample. In practise, the degree of squeezing, which can be
produced in atomic samples is small, such that the decorrelation
approximation is a reasonable assumption for the settings
considered in this article.
\bibitem{Error} Initially, $\langle J_x^2\rangle|_{t=0}=\langle
J_x\rangle^2|_{t=0}$. The time evolution of $\langle J_x^2\rangle$
is given by $d_t\langle
J_x^2\rangle=-2\left(\Gamma_{\text{cool}}+\Gamma_{\text{heat}}\right)\langle
J_x^2\rangle+\frac{N}{2}\left(\Gamma_{\text{cool}}+\Gamma_{\text{heat}}\right)
+N\left(\Gamma_{\text{cool}}-\Gamma_{\text{heat}}\right)\langle
J_x\rangle.$ Hence, $\langle J_x^2\rangle_{\infty}-\langle
J_x\rangle_{\infty}^2=\frac{N}{4}+\langle J_x\rangle_{\infty}$
(compare Eq.~(\ref{Main result})) in the steady state and the
error is by a factor $N$ smaller than $\langle J_x^2\rangle$ for
all times.
\end{thebibliography}
\end{document}